%% file: main.tex
\PassOptionsToPackage{table,xcdraw}{xcolor}
\documentclass[sigconf, nonacm, prologue, table, xcdraw]{acmart}
\newcommand\vldbdoi{XX.XX/XXX.XX}
\newcommand\vldbpages{XXX-XXX}
\newcommand\vldbvolume{19}
\newcommand\vldbissue{7}
\newcommand\vldbyear{2026}
\newcommand\vldbauthors{\authors}
\newcommand\vldbtitle{\shorttitle} 
\newcommand\vldbavailabilityurl{https://github.com/LeeBohyun/ZLeanStore}
\newcommand\vldbpagestyle{empty} 
\newif\ifextend
\extendtrue
\usepackage{algorithm}
\usepackage{makecell}
\usepackage[noend]{algpseudocode}
\usepackage{enumitem}
\usepackage{amsmath}
\usepackage{pifont}
\PassOptionsToPackage{capitalize,noabbrev,nameinlink}{cleveref}
\usepackage{cleveref}
\usepackage{subcaption}
\usepackage{multirow}
\long\def\comment#1{}
\def\module#1{\noindent\textbf{#1}}

\def\HLnote#1{\if@firstcolumn\reversemarginpar\else\normalmarginpar\fi\marginpar{\color{blue}#1}}
\newcommand{\marginnote}[1]{\if@firstcolumn\reversemarginpar\else\normalmarginpar\fi\marginpar{#1}}

\newcommand{\revision}[2][]{#2} 

\begin{document}
\pagestyle{plain}       

\title{How to Write to SSDs}
\ifextend
\subtitle{\large\color{gray}Extended version of a paper accepted to the Proceedings of the VLDB Endowment (PVLDB) 2026}
\fi
\author{Bohyun Lee}
\affiliation{%
  \institution{Technische Universität München}
  \postcode{43017-6221}
}
\email{bohyun.lee@tum.de}

\author{Tobias Ziegler}
\affiliation{%
  \institution{TigerBeetle}
  }
  \email{tobias@tigerbeetle.com}

\author{Viktor Leis}
\affiliation{%
  \institution{Technische Universität München}
}
\email{leis@in.tum.de}

\input{abstract}

\maketitle

\pagestyle{\vldbpagestyle}
\ifextend
\fi

\ifdefempty{\vldbavailabilityurl}{}{
\vspace{.3cm}
\begingroup\small\noindent\raggedright\textbf{PVLDB Artifact Availability:}\\
The source code, data, and/or other artifacts have been made available at \url{https://github.com/LeeBohyun/ZLeanStore}.
\endgroup
}

\input{intro}
\input{oop}
\input{comp}

\input{binpacking}
\input{deathtime}

\input{zns}
\input{gcgran}

\input{nowa}
\input{fdp}
\input{impl}

\input{eval}
\balance
\input{related}
\input{conclusion}

\begin{acks}
This work was funded by the Deutsche Forschungsgemeinschaft (DFG, German Research Foundation) -- 551650645.
The authors thank the Samsung Memory Research Center (SMRC), Yongho Song, and Meongchul Song for providing the infrastructure used for FDP experiments.
We acknowledge Western Digital and Matias Bjørling for providing the ZNS and comparison SSDs.
\ifextend
We thank the reviewers for their insightful questions. Selected responses are included in the next section as supplementary material.
\fi
\end{acks}


\ifextend
\input{appendix}
\fi

\begingroup
\setlength{\emergencystretch}{1.5em}
\Urlmuskip=0mu plus 1mu\relax
\bibliographystyle{ACM-Reference-Format}
\bibliography{ref}
\endgroup

\end{document}

%% file: abstract.tex
\begin{abstract}
This paper demonstrates that adopting out-of-place writes is essential for database systems to fully leverage SSD performance and extend SSD lifespan.
We propose a set of out-of-place optimizations that collectively reduce write amplification across both the DBMS and SSD layers.
We redesign the in-place, B-tree-based LeanStore to write out-of-place and support these optimizations, and evaluate it on diverse OLTP benchmarks, dataset sizes, and SSDs.
The final design improves throughput by 1.65--2.24$\times$ and reduces flash writes per operation by 6.2--9.8$\times$ on YCSB-A.
On TPC-C with 15{,}000 warehouses, throughput improves by 2.45$\times$ while flash writes decrease by 7.2$\times$.
Finally, we show that the architecture can seamlessly support novel SSD interfaces such as ZNS and FDP.
\end{abstract}

%% file: intro.tex
\section{Introduction}~\label{sec:intro}

\module{Modern DBMSs optimize for SSD characteristics.}
\revision[R3.W1]{
As SSD prices have continued to fall while DRAM costs stagnate, SSDs have replaced disks as the primary storage medium for high-performance DBMSs~\cite{DBLP:journals/pvldb/Leis24, scalestore, CFLRU, flashalloc, DBLP:conf/fast/ManeasMES22}.
Consequently, modern systems increasingly tailor their designs to SSD properties, 
such as internal parallelism~\cite{DBLP:journals/pvldb/HaasL23}, read-write asymmetry~\cite{DBLP:journals/pvldb/LeeAL23, WAR, DBLP:journals/pvldb/VohringerL23}, and SSD reads~\cite{readpriority, hidinglatency}.}

\module{Yet most DBMSs still do not optimize for SSD writes.}
However, few systems control database \emph{write behavior} to optimize for SSDs, even though SSDs handle writes very differently from hard disks. 
The overall SSD performance for out-of-memory workloads can be greatly impacted by how a database system writes to the device.
Furthermore, SSDs have a finite lifespan: each write wears out flash cells, unlike hard disks, which do not suffer from endurance limits~\cite{lerner2024principles, SSDhistory}.
However, despite these differences, most systems still issue writes as if targeting hard disks -- ultimately writing far more than intended while remaining unaware of the consequences.

\begin{figure}[t]
    \includegraphics[clip, width=\columnwidth]{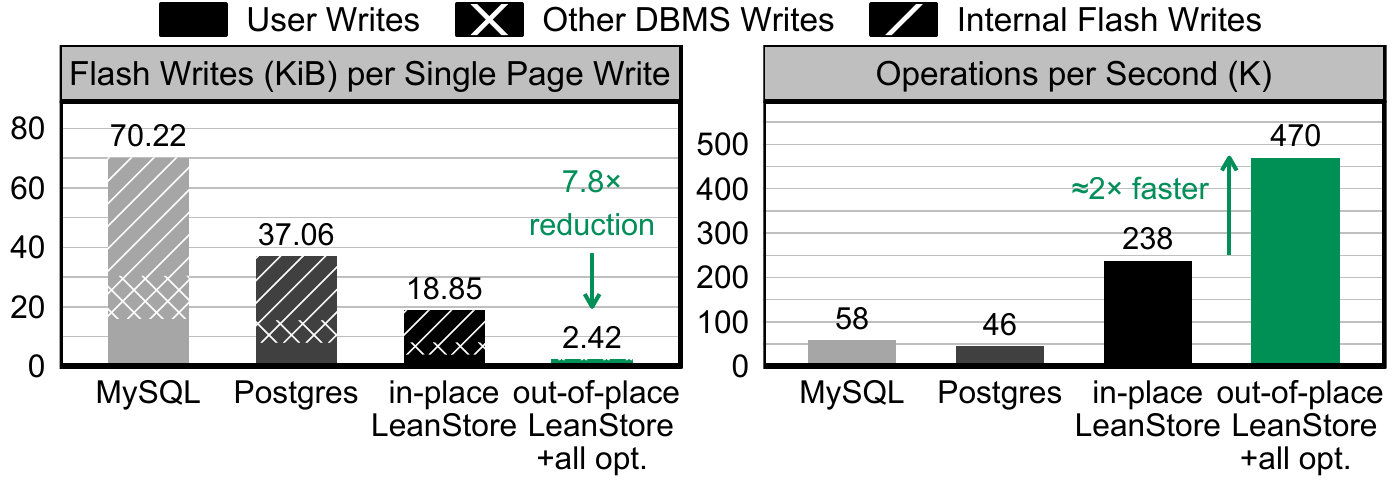}
    \caption{Flash writes per single page write and throughput (YCSB-A zipf $\theta$ = 0.8, Samsung PM9A3 SSD 90\% full, page sizes (KB): MySQL: 16, PostgreSQL: 8, LeanStore: 4)} 
    \label{fig:intro} 
\end{figure}

\module{Database systems write significantly more than expected.}
Once database writes accumulate beyond a certain threshold, ignoring SSD write characteristics becomes increasingly costly~\cite{ssdiq}.
We benchmark the B-tree-based engine LeanStore~\cite{DBLP:journals/pvldb/Leis24} on an enterprise SSD filled to 90\% using the YCSB-A workload (zipf $\theta$ = 0.8), 
running it until cumulative DB writes exceed four times the SSD’s capacity.
Surprisingly, as shown in the third bar of the first plot in \cref{fig:intro}, the in-place-write LeanStore baseline writes 18.85 KiB per 4 KiB B-tree node page write, 4.7$\times$ more than expected.
A similar trend appears in MySQL and PostgreSQL, as shown in the first two bars.

\module{Identifying the cause of write amplification.}
To investigate the source of the amplification, we categorize writes into three types (ignoring WAL writes because they are easy to handle):
(1) user writes (including evictions or checkpointing),
(2) additional DBMS-issued writes, and
(3) SSD-internal writes (obtained by OCP~\cite{ocp}). 
Here, we identify two main culprits behind this amplification.

\module{Culprit 1: DBMSs themselves write more.}
The DBMS often issues more writes than expected.
When comparing the user writes with the total amount of writes issued by the DBMS (user + other DB-issued writes) in \cref{fig:intro}, 
the latter is roughly \emph{double} the user writes due to the doublewrite buffering overhead.
This behavior is typical of in-place systems, including MySQL~\cite{mysql-dwb} and PostgreSQL (albeit indirectly)~\cite{postgresql/doc/dwb}, as shown in the first two bars in \cref{fig:intro}.

\module{Culprit 2: SSDs internally amplify DBMS writes.}
The SSD itself amplifies DBMS writes, a phenomenon known as SSD \emph{write amplification} (WA).
In our experiment, LeanStore experiences a Write Amplification Factor (WAF) of 2.36$\times$ (\cref{fig:intro}), meaning total flash writes are 2.36$\times$ the DBMS-issued writes.
This internal WA degrades performance over time and reduces SSD lifespan~\cite{lerner2024principles, arkiv/Dayan15}.
The Samsung PM9A3 data center SSD used in the experiment guarantees a lifetime of 5 years at 1 Drive Write Per Day (DWPD)---corresponding to an average write speed of only 11~MB/s.
LeanStore writes about 400~MB/s in the experiment, which means that the SSD would reach its endurance limit in about 1.5 months.

\revision[R2.D1]{
\module{DBMSs taking charge of minimizing end-to-end WA.}
A single page write is amplified first within the DBMS and then within the SSD, yielding a total WAF of DB~WAF~$\times$~SSD~WAF.
Conventionally, minimizing SSD WAF is delegated to lower layers---either the SSD or, indirectly, the filesystem~\cite{DBLP:conf/usenix/BjorlingAHRMGA21,DBLP:conf/fast/LeeSHC15,DBLP:journals/pvldb/KangCOL20}, 
while DB WAF is treated separately (e.g., LSM-tree WA optimizations~\cite{DBLP:journals/tos/LuPGAA17,rocksdb-saf}).
However, DBMS write issuance dictates both DB and SSD behavior; 
focusing only on DB WAF can paradoxically increase flash writes when SSD WAF is overlooked.
Thus, we argue that the DBMS, with the most workload knowledge at the highest layer, should coordinate writes with both DB- and SSD-level WA in mind.

\module{Reducing SSD writes with out-of-place techniques.}
To minimize flash writes per operation, we first move from in-place to out-of-place updates.
By removing the constraint of rewriting a page at a fixed location, the DBMS gains \emph{flexibility} in write placement. 
We then present several out-of-place-based optimizations that together minimize end-to-end WAF. }
These are implemented on LeanStore after converting to out-of-place writes. 
Using the same benchmark, the results (shown in the final bar of the two plots in \cref{fig:intro}) 
demonstrate that our out-of-place design with optimizations reduces flash writes by \emph{7.8$\times$} per operation compared to in-place LeanStore. 
This reduction is achieved by eliminating all other redundant writes, effectively reaching the near-optimal level of user writes alone.
Consequently, SSD lifespan is directly extended, while cost, power consumption, and carbon footprint per operation are reduced~\cite{carbon,DBLP:journals/usenix-login/Arpaci-Dusseau17}.
Throughput (operations per second, OPS) nearly doubles as well (right panel of \cref{fig:intro}).

\module{Contributions.}
\revision[R2.W1]{
The proposed optimizations systematically reduce cross-layer WA 
by considering how the write pattern shapes DB- and SSD-level amplification and their interplay.
Our design requires only modest changes to upper components of the storage engine, including the buffer pool, logging/recovery, and index structure, making it readily applicable to other B-tree-based systems.
The proposed optimizations are:}

\begin{itemize}[left=0.2cm]
\item \textbf{Compression \& page packing}:
\revision[R2.D4]{  
Page-wise compression reduces write volume, but with 4~KiB pages, its benefits vanish when compressed pages are misaligned or smaller than 4~KiB.
We introduce page packing, grouping compressed pages so each is read with a single 4~KiB access while retaining write and space savings.}

\item \textbf{Grouping by deathtime}:  
System-level out-of-place writes inherently require garbage collection (GC).  
To minimize WA during GC, we group pages by their expected invalidation time (``deathtime'') when placing them, derived from DB semantics.  

\item \textbf{ZNS support}:  
Our design naturally extends to Zoned Namespace (ZNS) SSDs, leveraging the existing DB GC and mapping.  
This enables full compatibility with ZNS, which inherently guarantees an (optimal) SSD WAF of 1.  

\item \textbf{Aligning DB and SSD GC units}:  
For non-ZNS SSDs, SSD GC-induced WA is mostly inevitable.
By estimating the physical placement inside SSDs, 
we find that the first key to mitigate SSD WAF is to align the DB GC unit with the SSD’s internal GC unit.  
When available, this is achieved using the FDP Reclaim Unit size; 
otherwise, it can be estimated using a ZNS-like pattern.

\item \textbf{NoWA pattern}:  
For commodity non-ZNS SSDs not featuring FDP, we introduce a NoWA (No Write Amplification) pattern that guarantees SSD WAF $~=$ 1, 
even at full device utilization.  
\revision[]{
\item \textbf{Using FDP placement hints}:  
For FDP-enabled SSDs, we show how to use placement hints that can replace the NoWA pattern.}
\end{itemize}
\noindent
\revision[R2.W1]{
The first two DB WAF optimizations are orthogonal yet complementary, 
while the remaining SSD-side optimizations may be chosen based on the underlying device.
Importantly, applying only the DB WAF optimizations without the SSD WAF ones can increase total WAF; 
the full set should therefore be viewed jointly.
In \cref{sec:oop}, we first motivate why the DBMS is best positioned to optimize SSD writes and why this necessitates out-of-place updates. 
We then present each optimization as structured in the following figure:
}
\begin{center}
\includegraphics[clip, width=0.95\columnwidth]{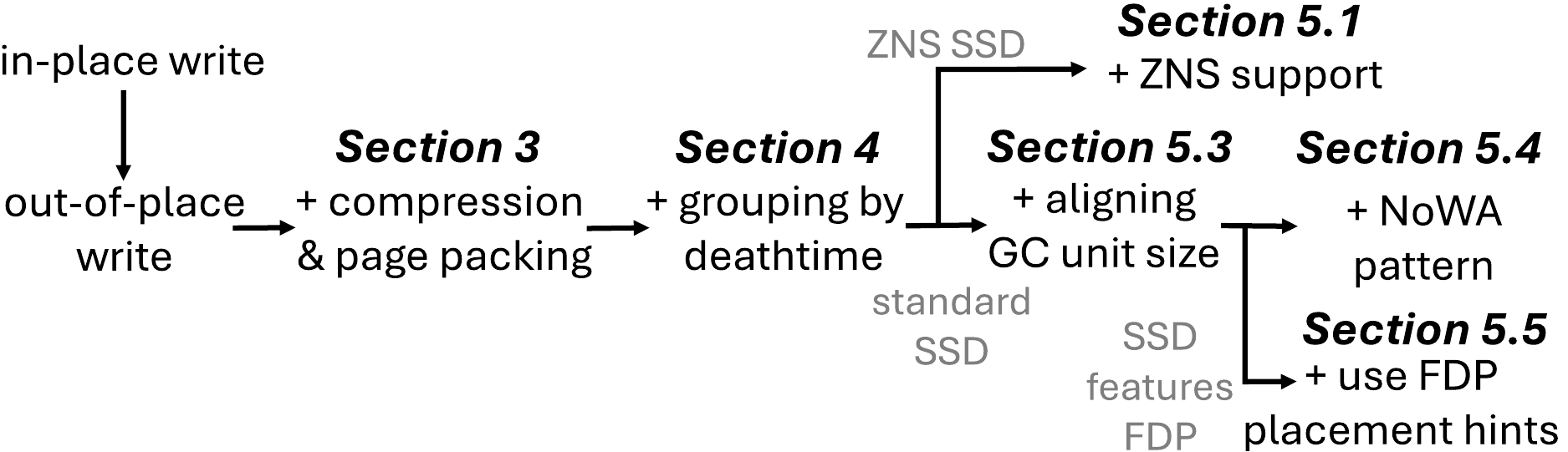}
\end{center}

%% file: oop.tex
\section{The Case for Out-of-Place Writes}\label{sec:oop}
We begin by reviewing SSD mechanics and WA, which influence both endurance and throughput.
We then argue that the DBMS is best positioned to manage end-to-end WAF at the highest layer.
To do so, however, requires adopting out-of-place writes, enabling the DBMS to use workload knowledge and SSD internals to reshape write patterns and thereby minimize flash writes per operation.

\module{Background: SSD endurance and write amplification.}
Flash SSDs differ from disks in two properties that matter for write-heavy workloads~\cite{DBLP:journals/pvldb/LeeAL23,flash-erase}.
First, flash cells only endure a limited number of program/erase cycles; each write gradually wears the cells, so device lifetime is tied to the \emph{volume of physical writes}~\cite{lerner2024principles,DBLP:conf/fast/YadgarYS15,dtp}.
Second, SSDs exhibit \emph{write amplification} because writes are performed at the page level (4-16~KiB), whereas erases reclaim entire blocks (many pages).
When a block containing still-valid pages must be erased to free space, 
the SSD first relocates those pages, increasing internal physical writes relative to host writes~\cite{DBLP:journals/pvldb/KakaraparthyPPK19,DBLP:journals/jsa/ImS10}.
Consider a DBMS issuing a write request for page $P$, as illustrated in the figure:
\begin{center}
  \includegraphics[clip, width=0.66\columnwidth]{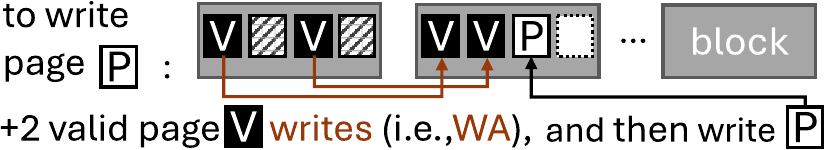}
\end{center}
To free space, the SSD’s garbage collector selects the first block in the figure as a victim; 
because two pages in that block remain valid, it copies them elsewhere before erasing the block.
Thus, one logical write of page $P$ results in three physical writes, directly reducing SSD lifespan and the bandwidth available to the host.

\module{DB WAF: User writes transformed by the DBMS.}
To understand more about the source of WAs in \cref{fig:intro}, we first examine the DBMS write path.
The DBMS persists a safe page copy before overwriting the target~\cite{mysql-dwb,postgresql/doc/dwb}, a mechanism known as \emph{doublewrite buffering} (DWB).
For this reason, in-place systems \revision[R2.D3]{such as MySQL, PostgreSQL, XtraDB, and CedarDB incur roughly $2\times$ the logical DB writes due to DWB~\cite{cedardb-dwb,xtradb-dwb,mariadb-dwb}}.
Without DWB, the system cannot recover from partial page writes.
Some SSDs expose limited atomic-write primitives~\cite{DuraSSD,DBLP:conf/fast/ZhengTQL13}, 
but typically only for 512- or 4{,}096-byte units and thus are not widely relied upon.

\begin{figure}
    \centering
    \begin{subfigure}[b]{0.325\linewidth}
        \centering
        \includegraphics[width=0.6\linewidth]{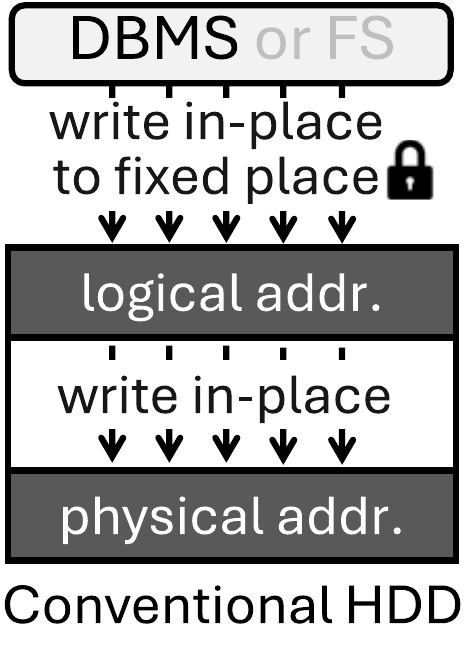}
        \caption{In-place write on a conventional disk}
        \label{fig:oop0}
    \end{subfigure}
    \hfill
    \begin{subfigure}[b]{0.325\linewidth}
        \centering
        \includegraphics[width=0.6\linewidth]{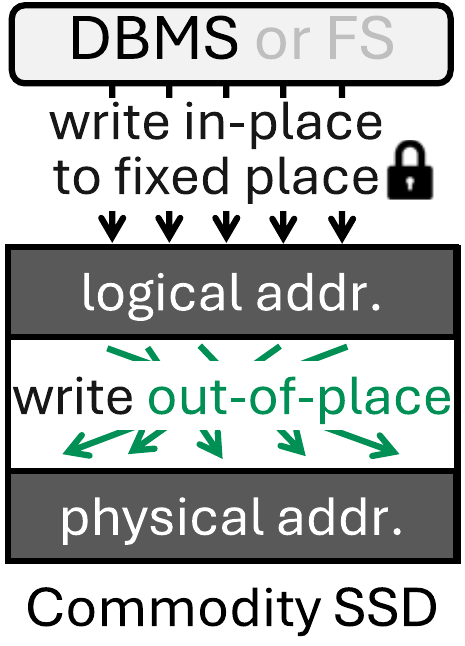}
        \caption{In-place write on a commodity SSD}
        \label{fig:oop1}
    \end{subfigure}
    \hfill
    \begin{subfigure}[b]{0.325\linewidth}
        \centering
        \includegraphics[width=0.6\linewidth]{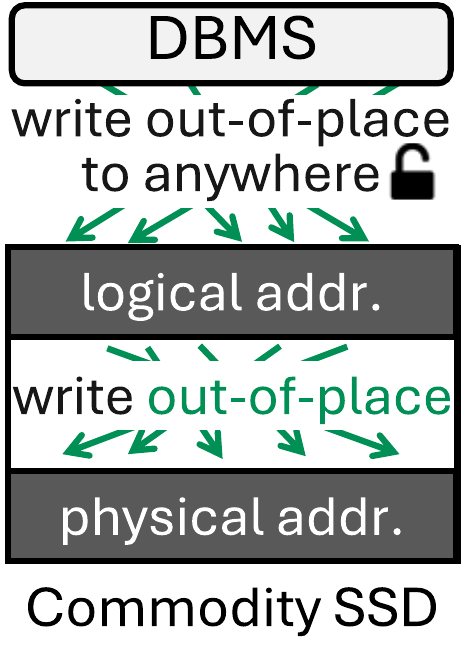}
        \caption{Out-of-place write on a commodity SSD}
        \label{fig:oop2}
    \end{subfigure}
    \caption{In-place vs. out-of-place writes on disk and SSD}
    \label{fig:oop}
\end{figure}

\module{SSD WAF: DBMS writes transformed by the SSD.}
Next, depending on the underlying block device, DBMS-issued writes are further amplified inside the device.
\revision[R2.D8]{%
Here, we assume DBMS is writing directly to the block device without a filesystem, 
which is common in modern systems~\cite{DBLP:conf/icde/NguyenL24,DBLP:conf/icit/HinesCA23,DBLP:journals/tos/AghayevWKNGA20, DBLP:journals/corr/SearsIG07,DBLP:journals/pvldb/GaffneyPBHKP22,DBLP:journals/pvldb/SkiadopoulosLKK21}. 
If a filesystem is used, however, it may alter the write pattern the device observes. }
On disks, logical and physical addresses mostly coincide (\cref{fig:oop0}); on SSDs, writes are inherently out-of-place (\cref{fig:oop1}), breaking this fixed mapping.
As a result, the physical consequences of any given write, particularly its internal placement and resulting WAF, can be severe and difficult to predict, except on specialized devices such as ZNS.
For instance, our prior study~\cite{ssdiq} (Figure~3) shows that several enterprise SSDs unexpectedly exhibit higher WAF (e.g.,~4) under even a simple hot/cold workload compared to a uniform-random workload.
This demonstrates that SSD WAF is not inherently managed by the device and can vary depending on workload patterns.

\module{Total WAF matters in the end.}
\revision[R2.W1]{ User writes are first amplified inside the DBMS and then inside the SSD.
To quantify this end-to-end amplification, we argue that systems should use \emph{total WAF} as a cross-layer metric.
Upon each page write for in-place LeanStore in \cref{fig:intro}, 
each logical write to persist a B-tree node is duplicated by DWB (\(\text{DB WAF} = 2.0\)) and further amplified within the SSD (\(\text{SSD WAF} = 2.36\)), resulting in the total amplification of 4.7.
Thus, workloads experience amplification at both the logical (DB) and physical (SSD) layers, whose combined effect is multiplicative: }
\[
\textit{Total WAF} = \textit{DB WAF} \times \textit{SSD WAF}
\]

\module{DBMS is best positioned to optimize for SSD writes.}
\revision[R2.W1]{
SSD WAF mitigation has mostly been attempted at lower layers (SSDs or filesystems)~\cite{DBLP:journals/pomacs/LangeNY25,DBLP:journals/cacm/LangeNY23,DBLP:journals/pvldb/KangCOL20,DBLP:conf/usenix/BjorlingAHRMGA21}, 
but these layers lack workload knowledge and merely receive host writes, leaving them intrinsically unable to change the write pattern that drives SSD WAF.
In contrast, \emph{the DBMS controls the write pattern}: by choosing what, when, and where to write (e.g., during eviction or checkpointing), it can proactively influence device behavior.
The DBMS possesses richer workload knowledge, including page hotness and access patterns, which are inaccessible to the SSD or filesystem.
Thus, with a working understanding of SSD internals, the DBMS can estimate amplification at both levels and adapt its write patterns accordingly. }

\module{Out-of-place writes are essential for minimizing total WAF.}
Even when a system aims to optimize SSD writes, the DBMS can do little if it performs in-place updates.
In-place systems fix each page’s file location (e.g., using the Page Identifier (PID) as an offset, as in \cref{fig:oop1}) and overwrite that location in the page size (16~KiB in InnoDB)~\cite{WObtree}.
Thus, the DBMS cannot alter write placement and therefore \emph{inherits} the device’s SSD WAF, forfeiting opportunities to optimize for SSDs.
Switching to out-of-place writes (\cref{fig:oop2}) unlocks flexibility, allowing the DBMS to group and place user writes as it chooses.
Out-of-place writes also eliminate doublewrite buffering: the old page remains valid until the new version is durably written, 
allowing crash recovery via the previous page and WAL replay.
Removing DWB nearly halves DB-issued bytes while preserving durability.

\module{Joint consideration of DB and SSD WAF is necessary.}
\revision[R2.W1]{To optimize total WAF, the DBMS must account for both DB and SSD WAF and their interplay.
Focusing on only one layer can counterintuitively worsen the other.
For example, an LSM-tree engine can lower DB WAF by increasing size tiering~\cite{spooky}, but doing so consumes more SSD space and can raise SSD WAF, ultimately worsening total WAF.
Thus, total WAF---not individual components---must guide optimization, reflecting the effects of DBMS writes at both layers. }

\module{Unlocking optimizations that minimize total WAF.}
In the following sections, we propose several out-of-place-based optimizations that, taken together, minimize total WAF.
At the DBMS level, Page-wise Compression (\cref{sec:comppagepack}) and Deathtime-Based GC (\cref{sec:GDT}) reduce \emph{DB WAF} by lowering logical write volume.
At the device level, we present techniques that ensure \emph{SSD WAF~=~1} across all three SSD types: ZNS (\cref{sec:zns}), standard (\cref{sec:nowa}), and standard FDP-enabled SSDs (\cref{sec:fdp}).
To our knowledge, this is the first demonstration of SSD WAF~=~1 on commodity SSDs---even at 100\% utilization and without fully sequential workloads.
Although DB-side and SSD-side techniques must be applied together to minimize total WAF, individual optimizations may be selectively adopted depending on the device (e.g., SSD or disk) and integration layer (e.g., filesystem).

%% file: comp.tex
\section{Reducing Writes with Compression}\label{sec:comppagepack}
Compression reduces DB WAF by lowering both the number of bytes written to storage and the overall dataset size.
Compressing pages before flushing to the device lowers the logical write volume.
However, if the characteristics of the underlying device are ignored, 
compression can backfire: misaligned or variable-length writes may increase read amplification or fail to reduce write and space at all.

\subsection{Page-Wise Compression}\label{sec:comp}
\module{How storage-level compression works.}
Assume a fixed database page size of 4~KiB and a buffer pool that caches at the same granularity.
Dirty pages are written to the storage device during eviction or checkpointing.
We compress each 4~KiB page \emph{independently} before persistence so it can later be read independently.
On reads, the page is decompressed before being placed in the buffer pool.

\module{Compression is difficult to apply with in-place writes.}
The SSD’s device-level write granularity complicates page-level compression for in-place systems.
\revision[R2.D4]{Compressing 4~KiB pages yields variable-length results (typically $<4$~KiB), but SSDs and filesystems commonly write in 4~KiB units.}
If we retain simple $PID \times 4\,$KiB addressing and overwrite in place, the device still writes a full 4~KiB block, eliminating any physical savings.
One could instead store compressed pages at variable offsets and maintain a PID$\rightarrow$offset mapping, but compressed sizes change as values are inserted or updated.
Offsets then drift, and relocating a page can cascade into shifting neighboring pages, making updates expensive.
Moreover, eviction follows recency rather than physical locality, so pages flushed together are rarely adjacent, hindering coalescing into aligned, sequential writes.
Consequently, many in-place DBMSs (e.g., PostgreSQL) forgo page-level compression~\cite{postgresql-toast}.
\revision[R2.D4]{Others offload compression to the filesystem (e.g., MySQL)~\cite{mysql-comp}.
However, this does not solve the fundamental issue with 4~KiB page sizes, as filesystems also issue I/Os in 4~KiB or larger units.}

\module{Optimization: \emph{easier} compression with out-of-place writes.}
\begingroup
\emergencystretch=3em
\sloppy
With out-of-place writes,\allowbreak managing variable-sized compressed pages becomes far more straightforward.
For each batch of pages, we compress individual pages, write the resulting compressed batch as a large sequential I/O, and record their new offsets and compressed sizes~\cite{wiredtiger-compaction}.
This avoids the challenges of in‑place systems, making compression far easier to implement.
\endgroup

\module{Compression ratio determines write reduction.}
How much DB WAF drops depends on data compressibility and the algorithm.
We load several datasets into LeanStore and compare compressed vs.\ uncompressed sizes for TPC‑C (1{,}000 warehouses), YCSB‑A (100~GB)~\cite{DBLP:conf/cloud/CooperSTRS10}, and real‑world datasets. 
Using LZ4~\cite{compression/lz4} and ZSTD~\cite{compression/zstd}, compression ratios (as \% of original; lower is better) range from 14\% to 49\%:

\smallskip
\centerline{%
\footnotesize
  \setlength\doublerulesep{0.5pt}%
  \setlength{\tabcolsep}{0.15cm}%
  \begin{tabular}{l | rrrrr}
    \toprule[1pt]\midrule[0.3pt]
    \textbf{Comp. Ratio (\%)} & \textbf{TPC-C} & \textbf{YCSB} & \textbf{Email} & \textbf{URL} & \textbf{Wiki} \\
    \midrule
    LZ4~\cite{compression/lz4} & 49.0 & 41.2 & 40.1 & 25.0 & 31.3 \\ 
    ZSTD~\cite{compression/zstd} & 36.5 & 34.4 & 27.4 & 14.5 & 17.8 \\ 
    \midrule[0.3pt]\bottomrule[1pt]
  \end{tabular}%
}
\smallskip

\noindent
These results confirm that most OLTP workloads will benefit from compression, as it reduces write volume and storage footprint.

\module{CPU overheads are negligible in most I/O-bound scenarios.}
Compression adds CPU work.
LZ4, for example, offers a good ratio at high speed~\cite{DBLP:journals/pacmmod/KuschewskiGNL24}, needing roughly 2--3 cycles/byte to compress and 0.5--1 cycle/byte to decompress~\cite{compression/lz4}. 
In most I/O‑bound or out‑of‑memory scenarios, the I/O savings dominate the CPU cost~\cite{DBLP:conf/sigmod/LeeM07}.
As we will show in \cref{sec:eval}, compression often improves overall performance, especially in OLTP settings.

%% file: binpacking.tex
\subsection{Page Packing After 4~KiB Compression}\label{sec:binpacking}

\begin{figure}
    \centering
    \begin{subfigure}[b]{0.3\linewidth}
        \centering
        \includegraphics[width=0.95\linewidth]{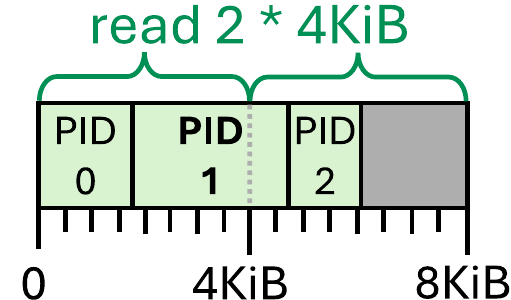}
        \caption{Read with \\ 4~KiB granularity}
        \label{fig:binpacking0}
    \end{subfigure}
    \hfill
    \begin{subfigure}[b]{0.3\linewidth}
        \centering
        \includegraphics[width=0.95\linewidth]{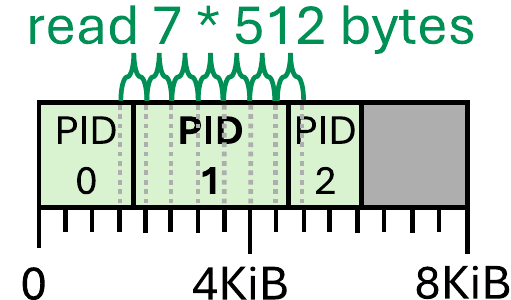}
        \caption{Read with \\ 512~B granularity}
        \label{fig:binpacking1}
    \end{subfigure}
    \hfill
    \begin{subfigure}[b]{0.3\linewidth}
        \centering
        \includegraphics[width=0.95\linewidth]{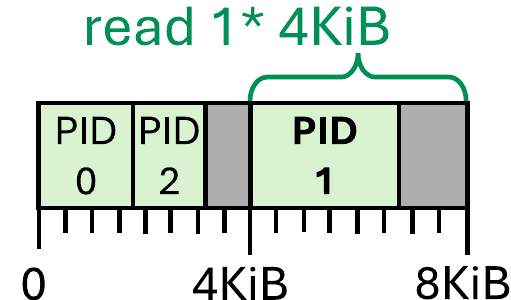}
        \caption{Read PID stored \\ by page packing}
        \label{fig:binpacking2}
    \end{subfigure}
    \caption{Total amount of reads when reading PID 1}
    \label{fig:binpacking}
\end{figure}

\module{Read is optimal with 4~KiB granularity and alignment.}  
Read latency strongly influences transaction latency~\cite{DBLP:journals/pacmmod/NguyenAZL25}. 
To identify the optimal read unit, we run an FIO microbenchmark~\cite{fio} on four enterprise SSDs: random reads from 512~B to 8~KiB in 512~B steps, aligned to request size. 
Each test ran for 30~minutes, with a single thread for latency (QD=1), and 32 threads (QD=64) for throughput.
Across all models, aligned 4~KiB requests achieved the lowest latency and highest throughput:
\begin{center}
  \includegraphics[clip, width=0.88\columnwidth]{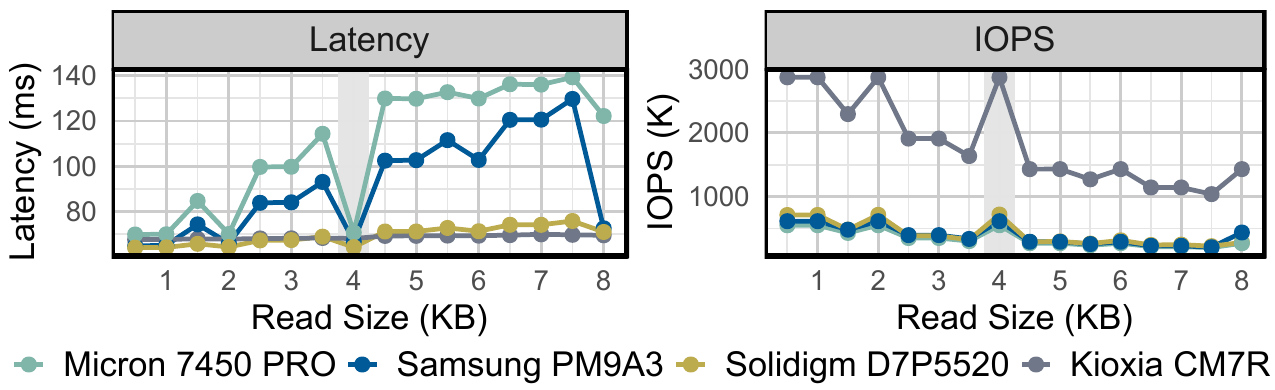}
\end{center}
\noindent
Smaller or misaligned requests, in contrast, incur additional internal reads~\cite{DBLP:journals/pvldb/HaasL23}.
Hence, 4~KiB is the most efficient size and alignment granularity for reading compressed pages.


\module{Read amplification can occur depending on the placement.}
Using 4~KiB pages with per-page compression yields variable-sized pages that, if poorly placed, can become misaligned on disk.
Such misalignment negates compression benefits by preserving write and space costs while introducing read amplification: a single PID read may trigger multiple physical reads~\cite{DBLP:journals/pvldb/HaasL23}.
For example, a 3{,}000~byte compressed page that crosses a 4~KiB boundary requires two 4~KiB reads, totaling $\approx2.73\times$ its size (\cref{fig:binpacking0}).
Even with 512~B reads, it still incurs $\approx1.19\times$ amplification (\cref{fig:binpacking1}).
This increases read latency and eliminates the intended compression gains.

\module{Optimization: 4~KiB-aligned page packing.}
To ensure each compressed page is retrievable with a single 4~KiB read, we use \emph{page packing}: align each compressed page to a 4~KiB boundary.
When writing a batch (e.g., 64 pages), we compress each page and place it into 4~KiB slots via a best‑fit binpacking algorithm~\cite{DBLP:journals/jal/Baker85}. 
Compressed pages never cross 4~KiB boundaries (\cref{fig:binpacking2}).
Compared to unaligned 4~KiB or 512~B reads, page packing lets each PID be fetched with one 4~KiB access---cutting I/Os relative to unaligned 4~KiB reads and remaining faster than 512‑byte reads.
In short, 4~KiB page packing yields fast, predictable reads with minimal amplification (with a small amount of bounded internal slack per page).

\begin{figure}
  \centering
  \includegraphics[clip, width=0.8\columnwidth]{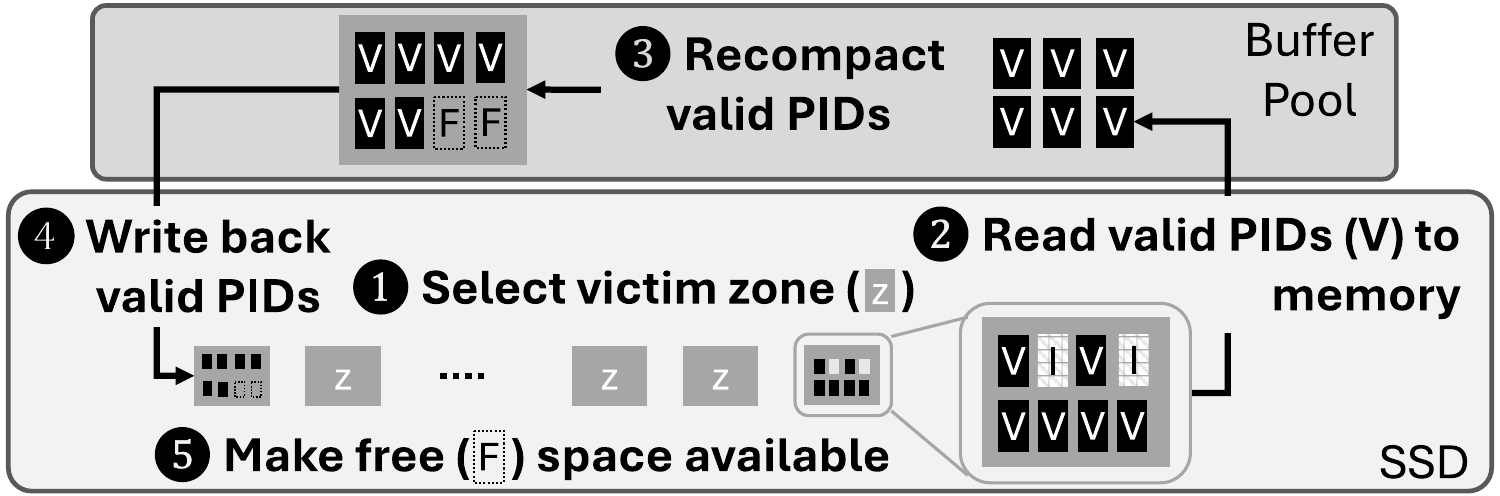}
  \caption{Garbage collection process}
  \label{fig:gc}
\end{figure}

%% file: deathtime.tex
\section{Deathtime-based GC}\label{sec:GDT}
Out‑of‑place systems require garbage collection (GC).
When a new version of a page is written, the previous version becomes invalid; 
stale data accumulates until capacity is exhausted.
We first explain how GC affects DB WAF, then introduce \emph{Grouping by Death Time} (GDT), 
which reduces WA for skewed workloads.

\subsection{Database-Level Write Amplification and GC}\label{sec:gc}

\module{GC overview.}
GC reclaims space occupied by stale pages, as illustrated in \cref{fig:gc}.  
Storage is divided into {\tt zones}, each containing multiple pages (eight in the figure), defining the GC granularity.  
When a new page is written, it is appended to a selected zone with available space, and the mapping table is updated while the old version is marked invalid.  
Over time, zones accumulate a mix of valid and invalid pages.
During GC, the garbage collector selects a \emph{victim zone} containing both valid and invalid pages (\ding{182}).  
In \cref{fig:gc}, 6 of 8 pages are valid, giving a 75\% valid ratio.
To reclaim space, valid pages not in memory are loaded (\ding{183}), compacted (\ding{184}), and rewritten (\ding{185}).
This copyback process reclaims space for two new writes (\ding{186}) but incurs WA, as six pages must be rewritten.
Garbage-collecting zones with more valid data therefore naturally yields higher WA.

\module{GC valid ratio drives DB WAF.}
The GC copyback drives DB WAF, as more pages are rewritten than originally requested.
WAF depends on the valid ratio of the victim zone in each GC cycle.
In \cref{fig:gc}, 75\% of the pages are valid, yielding a WAF of $1/(1-0.75)=4$, since 75\% of the data must be copied to free 25\% of the space.

\module{Skewed workloads require better GC.}
A common GC victim selection (\ding{182} in \cref{fig:gc}) is greedy; choosing the zone with the fewest valid pages.
This is optimal under uniform random workloads~\cite{DBLP:journals/cacm/LangeNY23, ssdiq}.
However, real-world workloads are typically skewed~\cite{DBLP:journals/tos/YadgarGJS21}, 
making greedy GC suboptimal~\cite{DBLP:journals/pvldb/KangCOL20, ssdiq}, as skew colocates hot and cold pages within zones, raising valid-page ratios and thus WA.

\module{DBMSs know better.}
To improve GC under skew, prior work proposes advanced GC and placement schemes---classifying data by hotness~\cite{DBLP:journals/pvldb/KangCOL20}, expected lifetime~\cite{dtp, DBLP:conf/fast/WangLLOSH22, lomet2020efficientlyreclaimingspacelog}, or based on spatial and temporal locality~\cite{DBLP:journals/pvldb/YuMKLMK22, DBLP:journals/pe/Houdt14, DBLP:conf/hotstorage/ShafaeiDF16}.
By grouping pages along these dimensions, these methods aim to reduce valid ratios in victim zones.
However, these techniques generally target SSD firmware or filesystems, which lack visibility into upper-layer workloads~\cite{Multistream, flashalloc, fdp}, limiting their ability to perform workload-aware GC.
In contrast, database systems possess richer workload and data semantics, making them well positioned to exploit intelligent GC schemes~\cite{DBLP:conf/sigmod/OzmenSSD10}.

\begin{figure}
    \centering
    \begin{subfigure}[t]{0.49\linewidth}
        \centering
        \includegraphics[width=0.95\linewidth]{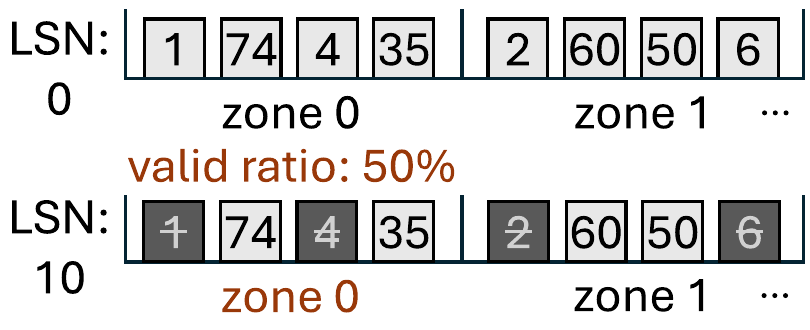}
        \caption{Random placement}
        \label{fig:beforegdt}
    \end{subfigure}
    \hfill
    \begin{subfigure}[t]{0.49\linewidth}
        \centering
        \includegraphics[width=0.95\linewidth]{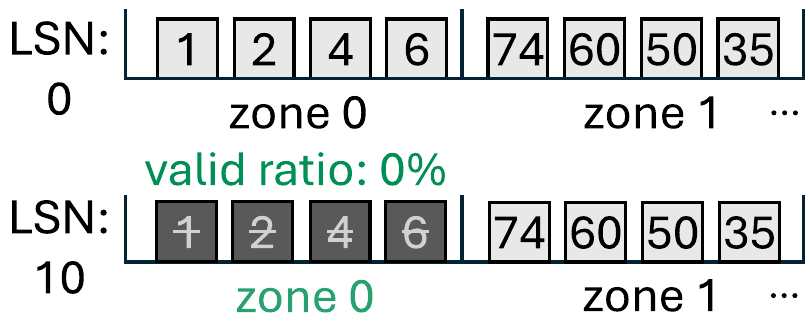}
        \caption{Placement using GDT}
         \label{fig:aftergdt}
    \end{subfigure}
    \caption{Impact of different placement strategies on WA\\
    ( {\raisebox{0.15em}{\setlength{\fboxsep}{0pt}\setlength{\fboxrule}{0.2pt}\framebox[0.8em][c]{\rule[-0.3em]{0pt}{0.8em}\scriptsize 1}}} denotes a page with deathtime 1.)}
    \label{fig:gdt}
\end{figure}

\subsection{Grouping Pages By Deathtime}
We introduce \emph{Grouping by Death Time} (GDT)~\cite{DBLP:journals/cacm/LangeNY23,DBLP:conf/eurosys/HeKAA17} into DB GC to reduce DB WAF.
GDT uses database semantics to estimate each page’s \emph{deathtime} (invalidation time).
This classification guides placement and GC, grouping pages that die together into the same zone and thereby reducing WA during reclamation (\ding{184}--\ding{186} in \cref{fig:gc}).

\module{Optimization: placing pages with similar deathtime per zone.}
GDT packs pages with similar deathtimes into the same zone so they are invalidated together. 
In \cref{fig:gdt}, eight pages with deathtimes 1, 2, 4, 6, 35, 50, 60, and 74 are written at LSN~0. 
If placed randomly (\cref{fig:beforegdt}), zones mix short- and long-lived pages, yielding a 50\% valid ratio at LSN~10 when greedy GC triggers. 
With GDT (\cref{fig:aftergdt}), pages are grouped by deathtime, so at LSN~10, victim zone~0 has no valid pages, eliminating WA for that cycle. 
Thus, aligning placement with deathtime substantially reduces GC-induced WA.

\module{Using write timestamps for deathtime estimation.}
To apply GDT, the system must estimate each page’s deathtime before writing it to SSD.
We extend each page header with a write timestamp, updated whenever the page is persisted~\cite{DBLP:journals/pvldb/VohringerL23}, and maintain a write history (WH) of the last $n$ timestamps (e.g., 4). GC writes do not update WH.
We estimate the next write time as: $\textit{EDT} = \textit{current\_lsn} + \frac{\textit{WH}_n - \textit{WH}_1}{n - 1}$,
extrapolating recent write intervals to predict when the page will next be overwritten.
For initial writes, pages are grouped by B-tree index ID to colocate pages with similar access patterns~\cite{DBLP:conf/sigmod/OzmenSSD10}.

\module{GDT-based placement strategy.}
When a batch of pages is written upon eviction, GDT first computes each page’s EDT.
Pages are then grouped and page-packed so that each group contains pages with similar deathtimes.
For each group, the strategy selects the zone whose existing pages have the closest average EDT; if no such zone is active, a new zone is opened.
This concentrates pages with similar deathtimes into the same zone.

\module{GDT-based GC.}
GDT placement is effective only when paired with GDT-aware GC; otherwise, GC writes and normal writes may intermix.
GC greedily selects victim zones until their cumulative invalid pages free the space of one zone.
All valid pages in these victims are loaded, sorted in descending EDT, grouped and packed, and written to an open zone with a matching EDT when possible; otherwise, they remain in their original zone.
Pages that are never rewritten (e.g., read-only pages) are assigned the maximum EDT to be treated as the coldest data.
This replacement during GC also compensates for initial EDT mispredictions.

\module{Reassigning EDT ranges after GC.}
After each GC cycle, zones may be full, partially full, or empty.
Empty and partially full zones are made available for normal writes.
The newly emptied zone is assigned the minimum EDT range, drawing short-lived pages; partially full zones naturally attract warm pages; and full zones retain the coldest pages.
This preserves aligned invalidation times within each zone, even after going through GC cycles.



%% file: zns.tex
\section{Minimizing SSD-Level WAF}~\label{sec:ssdwa} 

\module{SSDs also perform out-of-place writes.}
As discussed in \cref{sec:oop}, SSDs inherently use out-of-place writes and thus incur internal WA.
Along with the host write patterns, SSD WAF is shaped by the device’s internal architecture and algorithms~\cite{DBLP:conf/sigmod/LernerB21,lerner2024principles}.
Although SSD WAF is critical for performance and lifespan, it remains difficult to mitigate because SSDs operate as black boxes, exposing neither their internal behavior nor physical layout~\cite{ssdiq}.
Moreover, SSDs lack visibility into the workloads running above them.

\module{Different SSD interfaces for minimizing SSD WA.}
To bridge the knowledge gap between the host and SSD, recent work has explored more communicative SSD interfaces~\cite{DBLP:conf/sigmod/LernerB21}.
Zoned Namespace (ZNS) gives the host control over physical layout under certain write constraints, eliminating internal WA~\cite{DBLP:conf/hotstorage/Maheshwari21}.
Flexible Data Placement (FDP) allows the host to pass workload hints that the SSD can use to reduce internal WA~\cite{SSDhistory}.
While both interfaces are promising, their benefits depend on hardware support and proper integration with host-level management.

\module{Minimizing SSD WA regardless of SSD interfaces.}
We first show how our out-of-place write design naturally supports ZNS SSDs (\cref{sec:zns}).
When ZNS is unavailable and SSD-internal GC remains, 
we align DBMS and SSD GC units (\cref{sec:gcgranularity}) and introduce the \emph{NoWA write pattern}, which eliminates GC-induced WA even on commodity SSDs (\cref{sec:nowa}).
Finally, for FDP-enabled SSDs, we show how FDP placement hints can substitute for the NoWA pattern (\cref{sec:fdp}).

\subsection{Supporting Zoned Namespace SSDs}~\label{sec:zns}

\module{SSD WAF = 1 with ZNS.}
A key benefit of ZNS SSDs is that they eliminate internal WA, making total WAF (DB WAF $\times$ SSD WAF) effectively equal to the DB WAF, as GC occurs only at the DB level.  
To guarantee SSD WAF of 1, ZNS enforces sequential writes within each zone and shifts GC responsibility to the host.
This fits naturally with our design that already uses a translation layer and out-of-place writes with its own GC~\cite{DBLP:conf/usenix/BjorlingAHRMGA21}.  
With full workload knowledge, GDT-based GC combined with ZNS can perform GC more effectively than in-place systems on commodity SSDs (\cref{fig:oop1}).

\module{ZNS provides more space and reduces DB WA.}
ZNS SSDs expose more usable capacity because they require no internal GC and therefore no OP space~\cite{DBLP:conf/usenix/BjorlingAHRMGA21}.  
Commodity SSDs typically reserve 7--28\% of capacity for OP~\cite{DBLP:conf/fast/ManeasMES22}, whereas ZNS returns this space to the host~\cite{DBLP:conf/fast/YadgarYS15}.  
This extra capacity reduces DB WA: from the DB GC’s perspective, it acts as additional OP space, allowing GC to be delayed until more pages become invalid.  
When GC finally runs, zones have lower valid ratios, directly reducing DB WA~\cite{arkiv/Dayan15}.

\module{Optimization: compatibility with ZNS.}
To align our layout with ZNS, we set the database zone size (GC unit) to the ZNS zone size.  
Each zone maintains minimal metadata, including a write pointer and its state (empty, open, full).  
Writes are issued using zone-append commands to zone IDs.  
To reuse a full zone after GC copyback, the host issues a zone reset~\cite{DBLP:conf/usenix/BjorlingAHRMGA21}.  
Overall, our design integrates cleanly with ZNS and remains straightforward to implement.

%% file: gcgran.tex
\subsection{Why WA Persists on Commodity SSDs}\label{sec:ssdwaexists}

\revision[R2.W2]{\module{Standard SSDs can exhibit high WAF.}
When standard SSDs are used instead of ZNS, the same write pattern can incur high SSD WAF due to internal GC.
Although SSD-level WAF reduction has been widely studied, our prior work shows that modern enterprise SSDs still exhibit high WAF even under simple skewed workloads~\cite{ssdiq}.
Thus, even when DB WAF is reduced with our prior optimizations, excessive SSD WAF can negate overall gains.

\module{SSD WAF is shaped by DB GC behavior.}
For a fixed device, the dominant factor influencing SSD WAF is the write pattern generated by the DBMS~\cite{flashalloc}.
In our setting, DB GC shapes this pattern in three ways.
First, GC trigger points determine the SSD’s effective OP space: higher fill factors leave less OP space and increase SSD WAF. 
Second, DB GC granularity affects whether pages from the same zone remain colocated on SSD blocks.
Third, victim-zone selection determines which zones are rewritten together, influencing internal write clustering.
Together, these choices define the write stream the SSD receives and thus directly affect SSD WAF.

\module{DBMSs have more control than SSDs to mitigate SSD WAF.}
Although the write pattern is explicit at the DBMS layer, SSDs have no visibility into it~\cite{DBLP:journals/pomacs/LangeNY25}, making SSD-side mitigation difficult.
Prior work attempts to infer workload characteristics inside the SSD~\cite{DBLP:journals/pomacs/LangeNY25,DBLP:journals/cacm/LangeNY23,DBLP:journals/pvldb/KangCOL20}, but such inference is costly and inherently imperfect.
We take the opposite approach: the DBMS, which has full workload knowledge~\cite{SNIA_SDC_2025_TotalCostAndPerformanceOfSSDs}, guides writes to reduce SSD WAF directly.
By understanding SSD placement and GC behavior, the DBMS can \emph{proactively} structure writes to avoid GC-induced amplification.
}

\subsection{Aligning DB and SSD GC Granularity}\label{sec:gcgranularity}
A key insight available to the DBMS is that writes to the same zone share similar deathtimes.
This section explains why aligning DB and SSD GC granularity is the first step toward reducing SSD WAF using this insight.
We then show how the upper bound of SSD GC granularity can be inferred from ZNS-like patterns.
When an SSD features FDP, the Reclaim Unit size directly provides this value.

\module{Background: placement and GC granularity of SSDs.}
A superblock is the SSD’s primary append unit, formed by grouping erase blocks across planes and dies to exploit internal parallelism.
When a stream of page writes arrives, the SSD controller appends them to the active superblock and maps Logical Block Addresses (LBAs) to Physical Block Addresses (PBAs) after persistence.
Once the superblock fills, the SSD switches to a new one.
As the device fills and free space becomes scarce, GC reclaims space at a vendor-specific granularity, 
which may range from a single erase block to an entire superblock~\cite{SSDhistory,lerner2024principles}.

\begin{figure}[t]
    \centering
    \begin{minipage}[b]{0.49\linewidth}
        \centering
        \includegraphics[height=1.25in]{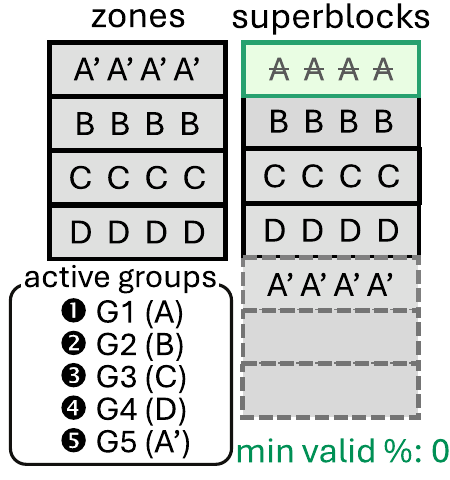}
        \caption{Zone size matches superblock size}
        \label{fig:gcunit0}
    \end{minipage}\hfill
    \begin{minipage}[b]{0.49\linewidth}
        \centering
        \includegraphics[height=1.25in]{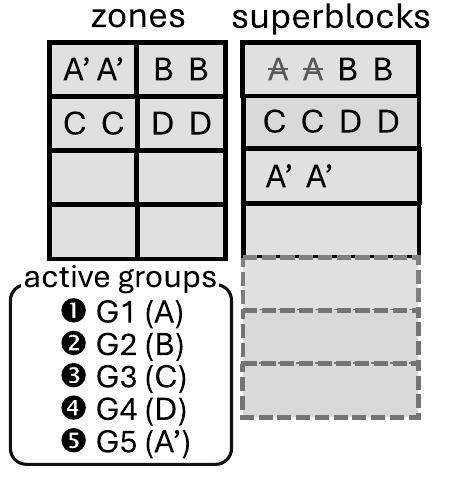}
        \caption{Zone size is smaller than superblock size}
        \label{fig:gcunit1}
    \end{minipage}
\end{figure}

\module{Estimating physical placement inside SSDs: simple case.}
To reason about physical placement, we use a simplified example that reflects general SSD behavior.
Suppose the SSD has seven superblocks, each holding four SSD pages (4~KiB), with three reserved as OP space, as shown in \cref{fig:gcunit0}.
The DBMS runs on top of the SSD, and DB GC is invoked when SSD capacity is nearly exhausted (four zones) and stops after reclaiming a target amount of space.
An \emph{SSD page} here refers to a 4~KiB LBA offset.
SSD pages belonging to the same DB zone share the same label.
For example, if zone~A spans LBAs 0--16~KiB, its first four SSD pages are labeled \emph{A}.
An \emph{active zone} is a zone currently receiving appends, and multiple active zones form an \emph{active group}.
A new zone becomes active only after all zones in the current group are fully written.
Thus, the number of zones the DBMS appends simultaneously determines the number of active zones and the size of the active group.

\module{Writes from the same zone share the same deathtime.}
\cref{fig:gcunit0} illustrates the case where only one zone is active and the zone size matches the superblock size.
When zone~A is appended (active group \ding{182}), its four writes (i.e., A pages) are placed together in the first superblock.
Later, when DB GC rewrites zone~A (active group \ding{186}), the new A' pages are written to the fifth superblock,
invalidating all old A pages in the first superblock and leaving it fully invalidated.

\module{DB GC granularity controls deathtime clustering.}
However, if the zone size is smaller than the superblock size, the first superblock does not become fully invalid.
In \cref{fig:gcunit1}, the zone size is half a superblock.
Even if the same active group is used as in \cref{fig:gcunit0}, SSD pages from zones A and B are interleaved within a superblock, mixing pages with different deathtimes~\cite{DBLP:conf/osdi/YangPGTS14}.
When DB GC rewrites zone~A (\ding{186}), only half of the superblock’s pages are invalidated.

\module{Aligning GC unit size to place writes with the same deathtime.}
These examples reveal two insights: (1) a write stream whose size matches the superblock is internally placed and GCed together, 
and (2) writes from a single zone form such a stream only when the DB and SSD GC units are aligned.
Thus, the SSD’s GC unit size determines the largest region whose pages can share the same deathtime and yield empty superblocks.
From the DBMS’s perspective, this maximum size can be used directly as the zone size.

\module{Optimization: using RU size (FDP-enabled SSDs).}
Standard SSDs usually do not expose the SSD GC granularity.
Fortunately, FDP explicitly exposes the SSD’s Reclaim Unit (RU), the smallest granularity at which SSD GC operates~\cite{SSDhistory}.
With this information, we can set the database’s zone size to match the RU size, ensuring that writes are aligned to the SSD’s internal GC boundaries.

\revision[R2.D7]{
\module{Optimization: using inferred GC unit size (standard SSDs).}
For SSDs without FDP, we approximate the upper bound of the SSD’s GC unit size using a ZNS-like write pattern.
With a single active zone, SSD WAF converges to~1 once the zone size meets or exceeds the internal append unit (\cref{fig:gcunit0}).
Thus, by gradually increasing the zone size and observing when WAF first reaches~1, we estimate this upper bound.

\module{Inferring the GC unit size with a ZNS-like pattern.}
We validate this approach on six enterprise SSDs, including two FDP-enabled devices, by varying the database zone size:
\begin{center}
  \includegraphics[width=0.8\linewidth]{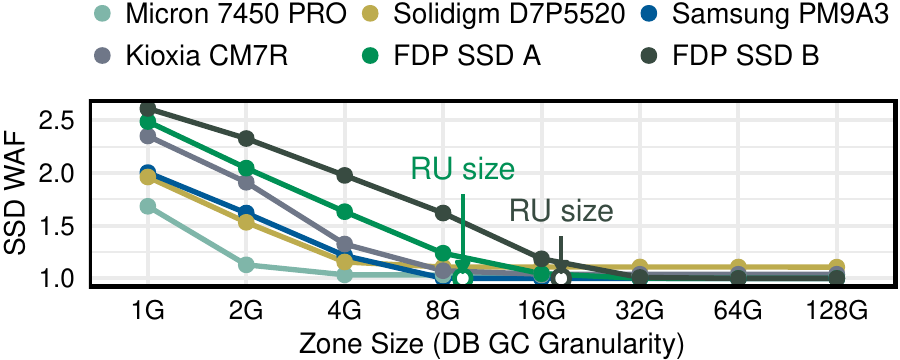}
\end{center}
\noindent
On FDP-enabled SSD~A, WAF drops to~1 at a 16\,GB zone size, consistent with its 8{,}496\,MB RU size lying between 8 and 16\,GB.
The same holds for FDP-enabled SSD~B, whose RU size is slightly above 16\,GB.
Across the other four enterprise SSDs, the inferred GC unit size typically falls between 4\,GB and 8\,GB.
Thus, the zone size at which WAF reaches~1 provides a practical upper bound on SSD GC granularity.
Although approximate, this bound is sufficient to avoid mixing DB writes with different deathtimes.
}
\revision[R3.D3]{For SSDs without OCP commands or FDP support, we recommend using a 32\,GB zone size as a safe upper bound.
Alternatively, one can indirectly estimate SSD WAF by checking whether throughput drops~\cite{ssdiq}. }


%% file: nowa.tex
\subsection{NoWA: SSD WAF=1 on Commodity SSDs}\label{sec:nowa}
With a single active zone, achieving SSD WAF~\(=1\) is straightforward (\cref{fig:gcunit0}).  
With multithreading, however, multiple zones are appended simultaneously, which complicates the problem.
In this section, we explain how WA arises from \emph{multiplexing} and \emph{frequency imbalance} among zones within the same active group~\cite{flashalloc}.  
We then introduce the NoWA write pattern, which prevents SSD-GC-induced WA and thereby guarantees SSD WAF~\(=1\) on commodity SSDs.  
For FDP-enabled SSDs, multiplexing can be avoided without NoWA; we discuss this in the next section.

\begin{figure}[tbp]
    \centering
    \begin{minipage}[b]{0.49\linewidth}
        \centering
        \includegraphics[height=1.25in]{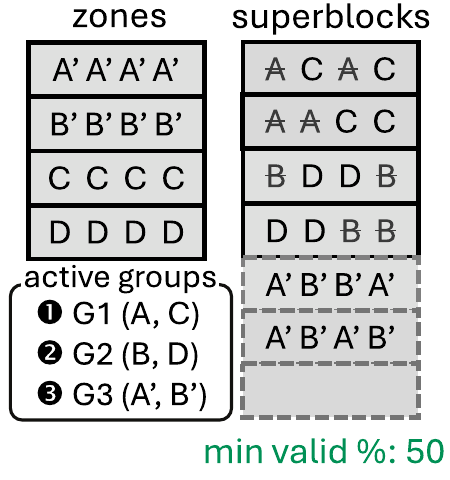}
        \caption{Imbalanced \mbox{fre-}\\\hspace{0.25em}quency in group \ding{182} and \ding{184}}
        \label{fig:multipleopen}
    \end{minipage}
    \hfill
    \begin{minipage}[b]{0.49\linewidth}
        \centering
        \includegraphics[height=1.25in]{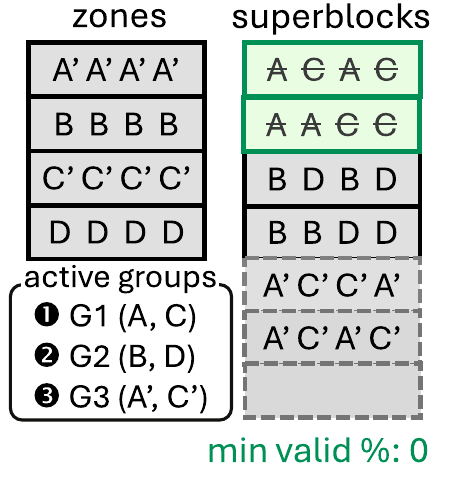}
        \caption{Balanced frequency in all active groups}
        \label{fig:noimbalance}
    \end{minipage}
\end{figure}

\module{Multiplexing occurs with multiple open zones.}
When multiple zones are appended concurrently, their write streams interleave across different superblocks~\cite{purandare2025valet}.
For example, if the DBMS writes to zones A and C at the same time (active group G1(A, C) in \cref{fig:multipleopen}), 
writes from both zones become mixed, scattering their data across two superblocks.
This multiplexing effect~\cite{flashalloc} arises because standard SSDs cannot distinguish writes from zones A and C.

\module{Frequency imbalance in an active group incurs WA.}
Multiplexing causes WA when zones in the same active group are invalidated at different rates.
In \cref{fig:multipleopen}, selecting zones A and B for DB GC (active group G3(A', B')) produces four partially invalidated superblocks that the SSD GC must later rewrite.
If, however, all zones in the active group are invalidated uniformly, the SSD can still produce superblocks with a 0\% valid-page ratio even under multiplexing.
For example, choosing G(A', C') in \cref{fig:noimbalance} fully invalidates two superblocks.
Thus, balanced invalidation within an active group eliminates SSD WA even when writes are multiplexed across zones.

\module{Optimization: NoWA pattern guarantees SSD WAF of 1.}
The \emph{NoWA pattern} eliminates SSD-GC--induced WA by ensuring that the SSD always has at least one fully invalidated superblock available when it reaches its minimum free--superblock threshold.
NoWA achieves this by estimating how pages are multiplexed within the SSD’s physical layout and enforcing two rules:
(1) defer opening new zones until all currently open zones are completely written, and
(2) detect and correct write--frequency imbalances among concurrently appended zones.
By selecting DB-GC victim zones in accordance with these principles, NoWA guarantees the existence of an empty superblock before SSD GC is triggered.
In effect, this write pattern preempts SSD-level GC.

\begin{figure}[tbp]
    \centering

    \begin{minipage}[b]{0.42\linewidth}
        \centering
        \includegraphics[height=1.25in]{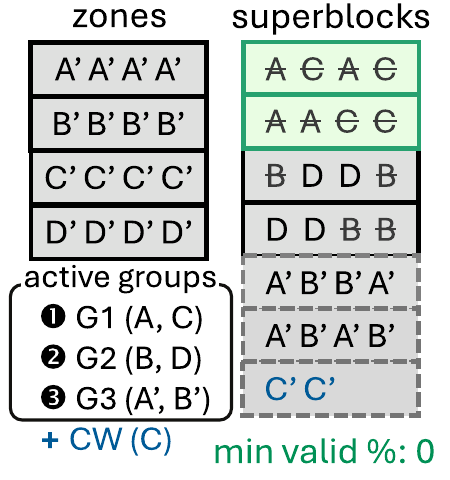}
        \caption{NoWA's compensation writes}
        \label{fig:nowa}
    \end{minipage}
    \hfill
    \begin{minipage}[b]{0.54\linewidth}
        \centering
        \includegraphics[height=1.25in]{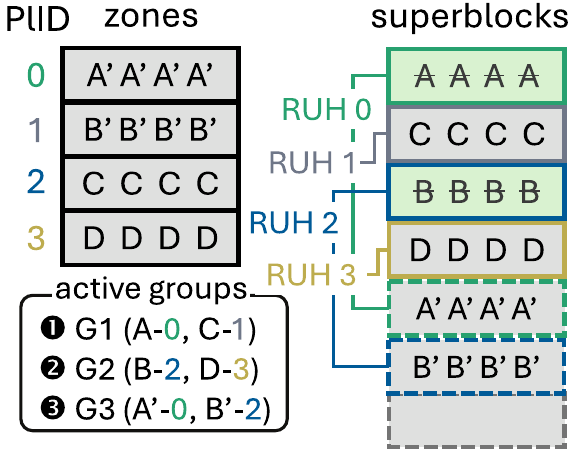}
        \caption{FDP placement hints avoid multiplexing}
        \label{fig:fdp}
    \end{minipage}
\end{figure}

\module{NoWA properties.}
The key idea behind NoWA is to eliminate frequency imbalance \emph{before} SSD GC is triggered.
As shown in \cref{fig:multipleopen}, if a new active group (e.g., G3(A', B')) creates uneven invalidation in earlier groups (e.g., A:2 vs.\ C:1 in G1),
the DBMS issues a \emph{compensation write} for the underrepresented zone (e.g., CW(C)), as illustrated in \cref{fig:nowa}.
This compensation must occur before the SSD reaches its minimum free--space threshold, which can be estimated using SSD--iq~\cite{ssdiq}.
The compensation write recompacts the scattered pages of zone~C into a single superblock, invalidating their older versions and effectively producing two fully invalidated superblocks.
Thus, NoWA proactively restores balance whenever imbalance occurs, ensuring that subsequent SSD GC cycles operate on fully invalidated superblocks.

\module{Compensation writes are not entirely redundant.}
Compensation writes shift a small portion of WA from the SSD to DBMS-level GC.
Although they increase DB WAF slightly, the overhead is modest because the DBMS controls zone selection.
Before selecting a victim zone, GC can check whether issuing a compensation write would raise valid-page ratios in later rounds; if so, it simply chooses another zone.
Overall, this small increase in DB WAF is far outweighed by the reduction in SSD WAF (e.g., from 2.3 for in-place to 1 with out-of-place LeanStore using NoWA; see \cref{fig:intro}).

\module{Flexible zone sizing with NoWA.}
NoWA also allows flexible configuration of DB zone size by adjusting the maximum number of concurrently open zones.
NoWA holds as long as $\mathrm{max\ open\ zones} \times \mathrm{zone\ size}$
equals or is a multiple of the inferred SSD GC unit.
For example, on FDP-enabled SSD~A, 16 active zones of 513~MB each, or 8 zones of 1{,}026~MB, satisfy this requirement.
This flexibility enables smaller zones to reduce GC-induced tail latency while preserving NoWA’s guarantee of SSD WAF~=~1.

\module{NoWA is worthwhile, even if it is imperfect.}
\revision[R2.D11]{NoWA may not always completely eliminate SSD WA, 
since SSDs may reorder or relocate data depending on scheduling, load, flash chip state, open-superblock limits, or wear-leveling~\cite{lerner2024principles}. 
However, it still provides substantial practical benefit: even partial mitigation of page mixing reduces WA.
Nevertheless, NoWA achieves SSD WAF~=~1 on six enterprise SSDs from diverse vendors and capacities (\cref{fig:varyingssd}).}

%% file: fdp.tex
\revision[R2.D9]{
\subsection{Using FDP Placement Hints}\label{sec:fdp}
\module{FDP exposes RU handles for host-side placement hints.}
When an SSD supports FDP, the NoWA pattern becomes unnecessary to achieve an SSD WAF of 1.
In addition to the RU size, FDP-enabled SSDs expose \emph{Reclaim Unit Handles (RUHs)}, each managing a disjoint set of RUs~\cite{fdp}.
The number of RUHs determines how many independent write streams the SSD can maintain without multiplexing.
The host can leverage this by providing placement hints when issuing writes, preventing pages with different deathtimes from being mixed internally~\cite{song_fast26_warp}.
In our design, this directly corresponds to the maximum number of active zones.

\module{Optimization: using FDP to avoid multiplexing.}
FDP placement hints are straightforward to use in our design.
We assign each zone a placement ID (PlID) modulo the number of RUHs and persist this mapping per zone.
For example, with four RUHs (PlIDs 0--3), zones A--D map to PlIDs 0--3, so writes to zone A always use PlID~0.
Under the write pattern shown in \cref{fig:multipleopen}, writes to different zones are directed to different RUHs, preventing interleaving between zones such as A and C (\cref{fig:fdp}).
With the internal SSD information exposed by FDP, an SSD WAF of~1 is achievable as long as the zone size equals the RU size and the number of open zones does not exceed the RUH count (as shown in \cref{tab:fdpresults}). }

%% file: impl.tex
\section{Implementation}\label{sec:impl}

\module{From in-place to out-of-place writes.}
To evaluate our techniques, we extend LeanStore~\cite{DBLP:journals/pvldb/Leis24}, a B-tree system that originally performs in-place writes.
We integrate vmcache~\cite{DBLP:journals/pacmmod/LeisA0L023} as the buffer pool to support 1:N PID-offset mappings, 
enabling out-of-place writes.
The resulting prototype, \emph{Z (Zoned) LeanStore}, shows that our optimizations apply to an arbitrary page-based system.
\revision[R2.W1]{We first outline the write path to illustrate how the optimizations work together.
We then describe necessary extensions to other components~\cite{WObtree, WiredTiger_Checkpoint_Arch} to implement our techniques without compromising durability. }

\subsection{Design Overview}

\module{How the main components support optimizations.}
We modify four DBMS components: the buffer manager, I/O interface, space manager, and garbage collector.
The buffer manager computes each page’s EDT using write timestamps stored in the page header.
The I/O interface performs compression, page packing, and selects the backend based on the SSD type (e.g., ZNS or FDP).
Storage space is partitioned into zones, sized according to the inferred SSD GC unit.
The space manager maintains PID-offset and reverse mappings, tracks zone metadata, and selects zones for placement based on GDT.
The GC reclaims space in both foreground and background, choosing victims based on GDT while preserving the NoWA pattern.

\begin{figure}[t]
    \centering
    \includegraphics[clip, width=\linewidth]{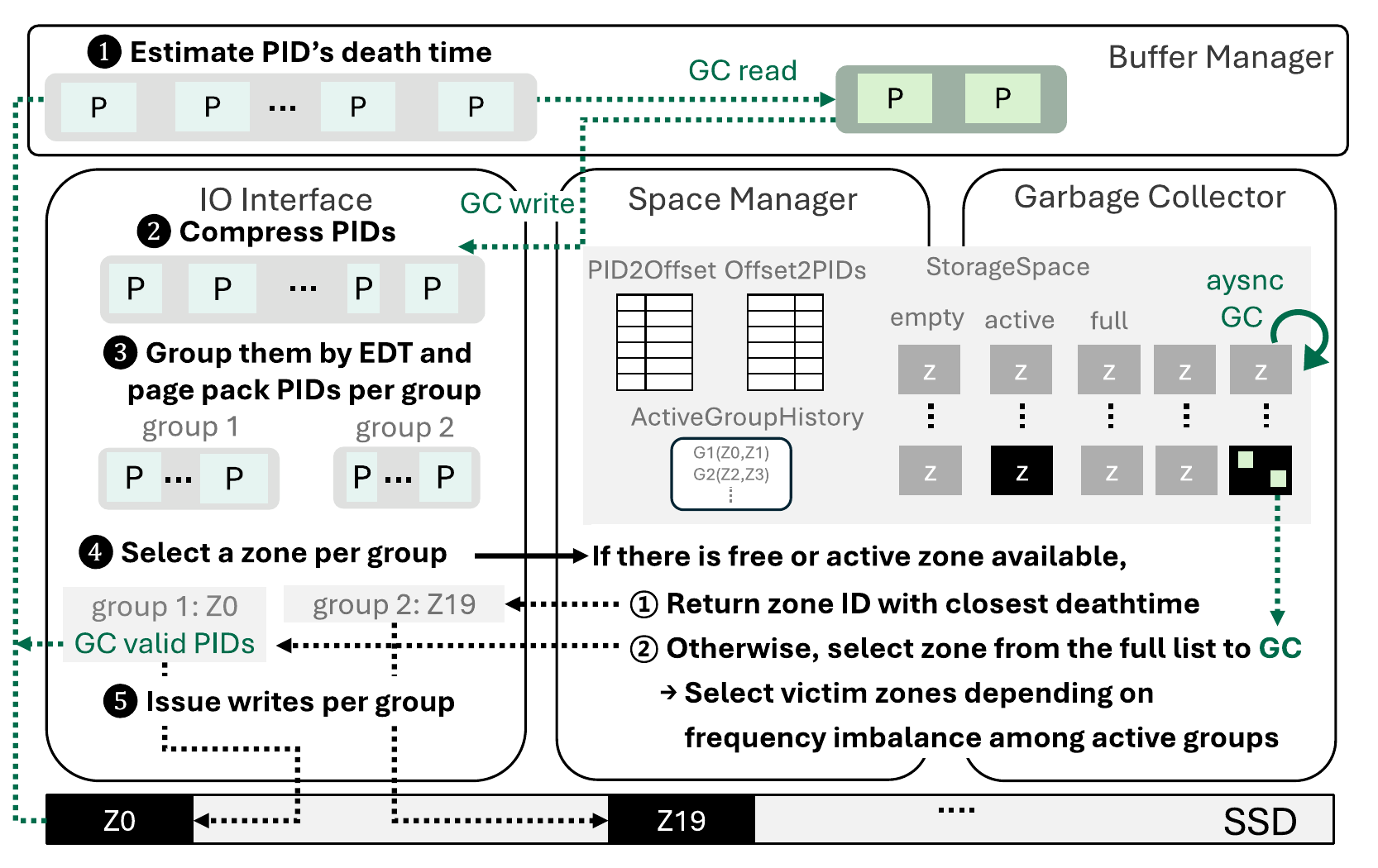}
    \caption{Overview of the write path with all optimizations}
    \label{fig:writepath}
\end{figure}

\module{Write path.}
Write requests follow the path illustrated in \cref{fig:writepath}.
Suppose a batch of PIDs must be flushed upon eviction.
First, each PID’s EDT is computed from previous write LSNs stored in the page header.
The PID, its contents, and its EDT are passed to the I/O interface (\ding{182}), which compresses pages (\ding{183}) and groups PIDs with similar EDTs (\ding{184}).
The space manager then assigns a target zone for each group (\ding{185}): 
if active zones exist, the one whose average EDT is closest is chosen (\textcircled{1}); 
otherwise, a new zone is opened from the empty or full list, respecting the active-zone limit.
If a zone is full, GC is triggered (\textcircled{2}).
During GC, valid pages not in the buffer pool are read in and rewritten following the same write path (\ding{182}-\ding{184}).
When multiple zones are reclaimed, valid PIDs are sorted by descending EDT and written into zones with the highest invalidation counts, placing the coldest pages into the coldest zones.
After GC completes, the originally requested write proceeds (\ding{186}).


\subsection{Implementation Details}

\module{Sharing the buffer pool as a cache for GC.}
The GC shares the buffer pool with worker threads to perform GC I/Os, avoiding memory partitioning and allowing full utilization when GC is inactive.
This reduces disk reads during GC as some valid PIDs may already be in memory.
Although GC may harm the hit ratio, it often acts as a prefetcher, as GDT-based GC tends to load PIDs that will be accessed soon, thus reducing its negative impact.

\module{Securing clean buffer frames for GC reads.}
A full buffer pool can create contention between eviction and GC, as both require frames.
To avoid eviction writes during GC, only clean frames may be reused for GC reads.
Thus, the system maintains a reserve of clean pages, integrated with fuzzy checkpointing~\cite{leanstore-logging}.
If the reserve falls below the threshold, the checkpointer flushes pages during non-GC phases.
In practice, this mechanism is rarely invoked because checkpointing already keeps enough clean frames available.

\module{I/O buffer for compression and page packing.}
The I/O interface uses a per-thread buffer to issue I/O requests.
PIDs are compressed with LZ4~\cite{compression/lz4}, grouped by GDT, page-packed, and issued as an I/O request.
For reads, a 4~KiB-aligned buffer fetches the data; after decompression, the requested PID is inserted into the buffer pool.

\module{Space metadata.}
The space manager maintains three types of metadata and updates them whenever a write occurs.
First, \texttt{PID2Offset\-Table} maps each PID to its on-disk offset and compressed size.
Second, \texttt{Storage\-Space} provides the reverse mapping.
It tracks which PIDs occupy each 4~KiB offset and stores zone metadata such as fill factor, page and zone invalidation counts, average EDT, and state.
Finally, to implement the NoWA pattern, \texttt{ActiveGroupHistory} records the active group each time a new set of zones is opened.

\module{Supporting various GC strategies.}
The garbage collector activates whenever active zones nearly run out of free device space.
Depending on the enabled configuration, it runs victim selection strategies such as greedy, GDT-based GC, and/or NoWA-aware GC.
Thus, the user can run GC that fits the workload characteristics.

\revision[R2.W2]{
\module{Compatibility with ZNS SSDs and FDP.}
The I/O layer detects the device type and selects the appropriate backend via \texttt{ioctl}.
Standard SSDs use an \texttt{io\_uring}-based backend.
For ZNS SSDs, writes use zone-append operations, zones are reset after GC, and the system tracks the device’s limit on open zones.
For FDP-enabled SSDs, we query the RU size and number of RU handles, set the maximum number of active zones accordingly, and apply placement hints by mapping zone IDs modulo the number of handles.
}

\revision[R2.W1, R2.D1]{
\module{Logging and recovery.}
\begingroup
\emergencystretch=3em
\sloppy
We use per-thread logs with continuous checkpointing~\cite{leanstore-logging}.
To ensure recoverability, the system logs \texttt{PID2Offset} updates.
We enforce write-ahead rules: page data is persisted before its mapping update is committed.
Each checkpoint persists snapshots of the \texttt{PID2Offset\-Table} and the \texttt{Active\-Group\-History}.
Recovery is performed by first locating the checkpoint LSN, reloading these snapshots, and replaying subsequent WAL entries; this naturally also reconstructs metadata.
Finally, the \texttt{Storage\-Space}\allowbreak layout is rebuilt from the final \texttt{PID2Offset\-Table}.
\endgroup
}

\begin{figure*}[t]
    \centering
    \begin{subfigure}{0.23\textwidth}
        \centering
        \includegraphics[width=\linewidth]{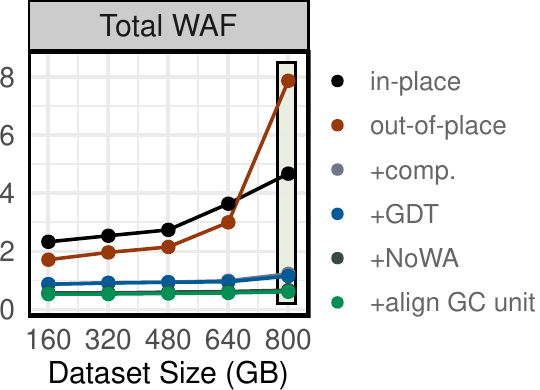}
        \caption{Total WAF}
        \label{fig:drilldown:a}
    \end{subfigure}
    \hfill
    \begin{subfigure}{0.34\textwidth}
        \centering
        \includegraphics[width=\linewidth]{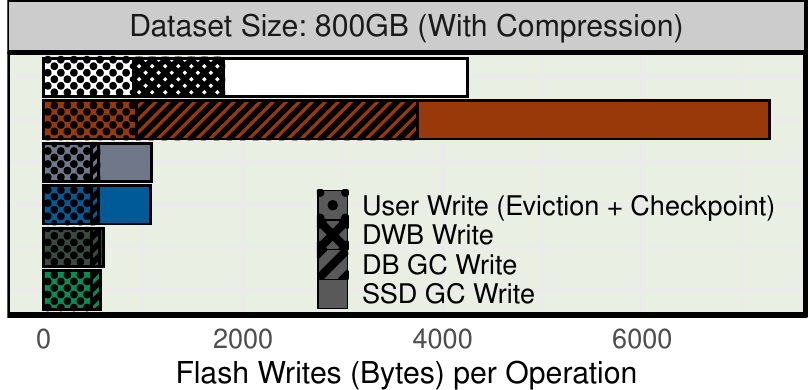}
        \caption{Write drilldown with compression}
        \label{fig:drilldown:b}
    \end{subfigure}
    \hfill
    \begin{subfigure}{0.42\textwidth}
        \centering
        \includegraphics[width=\linewidth]{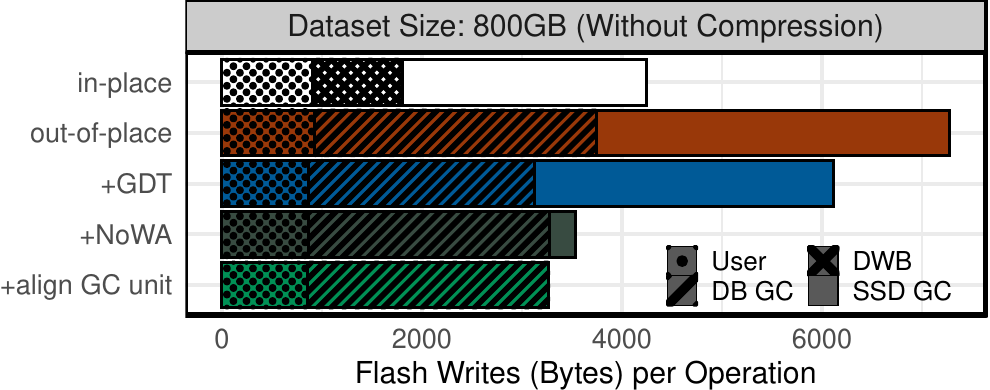}
        \caption{Write drilldown without compression}
        \label{fig:drilldown:c}
    \end{subfigure}
    
    \caption{Total WAF on different dataset sizes and write drilldown with 800~GB dataset (YCSB-A zipf $\theta$ = 0.8, Samsung PM9A3)}
    \label{fig:drilldown}
\end{figure*}

%% file: eval.tex
\section{Evaluation}\label{sec:eval}
Our evaluation shows that the proposed out-of-place optimizations substantially reduce total WAF while improving throughput and SSD lifespan 
across various devices (with larger gains on ZNS or FDP), dataset sizes, and benchmarks.  
We focus primarily on write-related metrics, as minimizing flash writes is our main goal.

\subsection{Setup and Methodology}
\module{Hardware setup.}
We evaluate on eight enterprise SSDs from five vendors.
Experiments run on two servers: one with 360~GB DRAM and a 96-core AMD EPYC~9654P, and another with 500~GB DRAM and a 64-core Intel Xeon Gold~6342, both running Ubuntu~24.04~LTS.
The first is used for general tests with 64 worker threads; the second for FDP-enabled SSDs with 32 worker threads.

\module{OLTP benchmarks and DBMS configuration.}
We use two OLTP benchmarks.  
YCSB-A~\cite{DBLP:conf/cloud/CooperSTRS10} (50\% reads, 50\% updates on a fixed-size dataset) is used for performance analysis, and TPC-C is used to assess generality and behavior under growing workloads.  
Both benchmarks target write-intensive OLTP scenarios that become I/O-bound once the dataset exceeds memory.  
\revision[R2.D10]{To reflect realistic OLTP deployments, we set the buffer pool to 5--20\% of the dataset to cache the effective working set~\cite{nvppl,Gray1993DebitCredit,gray1987fiveminuterule}.}  
The WAL size matches the buffer pool unless stated otherwise.

\module{Evaluation method.}
Before each run, we issue \texttt{blkdiscard} to reset SSD mapping state.  
Benchmarks run until cumulative writes reach at least 4× the device capacity to ensure steady state~\cite{ssdiq}.  
We report average throughput, DB WAF, SSD WAF, buffer hit ratio, and CPU/memory usage over the final hour, after configurations have issued comparable writes.  
DB WAF is total DB-issued writes divided by eviction or checkpoint writes, and SSD WAF is total physical writes (via OCP~\cite{ocp}, when supported) divided by DB-issued writes.

\subsection{Main Performance Result}\label{sec:mainperfresult}
To analyze the impact of each optimization on total WAF, we run YCSB-A (skew = 0.8) across five dataset sizes on a Samsung PM9A3 SSD (894~GB), 
incrementally enabling our optimizations.  
Before introducing GDT, random placement and greedy GC are used.  
When out-of-place writes are enabled, up to 16 open zones are allowed.
The zone size is set to 250~MB prior to aligning the GC unit and to 512~MB afterward, following the configuration inferred from the ZNS-like workload in \cref{sec:gcgranularity} (16 $\times$ 512~MiB = 8~GB in total).
\module{Total WAF.}
As shown in \cref{fig:drilldown:a}, each optimization reduces total WAF across all dataset sizes.
The 160~GB dataset sees a 4.4× reduction (2.33→0.53), while the 800~GB dataset reaches a 7.8× reduction (4.72→0.60) with all optimizations enabled.
Compression contributes the largest improvement by sharply lowering DB WAF.
The NoWA pattern provides the second-largest reduction, with its benefits increasing for larger datasets.

\module{Write drilldown (with compression).}
\cref{fig:drilldown:b} breaks down writes for the 800~GB dataset into user, DWB, DB GC, and SSD GC writes.
In the in-place baseline, DWB dominates, accounting for half of all DB-issued writes.
Switching to out-of-place writes removes DWB but makes DB GC the primary overhead, even increasing total WAF by 1.66×.
Compression with GDT reduces DB GC writes, though SSD GC remains.
Finally, the NoWA pattern together with aligned GC units eliminates SSD GC entirely, leaving user writes dominant and driving total WAF near the compression ratio.

\module{Write drilldown (without compression).}
Compression shrinks the 800~GB dataset to 418~GB, expanding OP space to 53\% of the SSD and reducing the GC valid ratio from 75\% to 14\% (\cref{fig:drilldown:b}).  
This lowers DB WAF and masks other effects, so we omit compression and page packing to isolate the remaining techniques in order of application.  
As shown in \cref{fig:drilldown:c}, out-of-place writes with all optimizations improve total WAF by 31\% over in-place and 2.18× over naïve out-of-place.  
GDT reduces DB WAF from 4.02 to 3.36, while NoWA increases it to 3.58 due to compensation writes; the resulting drop in SSD WAF outweighs this increase.  
GDT mitigates DB WAF, while SSD-side optimizations (NoWA and aligning the GC unit) become more impactful without compression.  
\cref{fig:dbsz} shows that total WAF improves by at least 2× for smaller datasets.

\module{Compression remains highly beneficial.}
Although the other optimizations are effective, compression remains highly beneficial and complements them well.  
Compression provides additional OP space for DB GC, lowering valid ratios by widening the invalidation window~\cite{ssdiq}.  
It also improves deathtime estimation by allowing more writes before GC triggers, giving a larger timestamp window.  
Together, these effects reduce DB WAF during GC.  
Compression also reduces space amplification and often improves throughput, making it a recommended choice when datasets are compressible.

\subsection{Performance Drilldown}~\label{sec:drilldown}
To examine how optimizations affect other metrics, 
\cref{tab:drilldown} summarizes the 800~GB experiments in \cref{fig:drilldown}.

\begin{table}[t]
\setlength\doublerulesep{0.5pt}
\setlength{\tabcolsep}{0.06cm} 
\small
\centering
\captionsetup{skip=6pt}
\caption{Performance drilldown per proposed optimization (YCSB-A zipf $\theta$ = 0.8, DB size: 800~GB, Samsung PM9A3)}
\label{tab:drilldown}
\begin{tabular}{l | r r r r r r} 
\toprule[1pt]\midrule[0.3pt]
\textbf{Version} & \textbf{OPS} & \textbf{DB} & \textbf{SSD} & \textbf{Logical W.} & \textbf{Physical W.} &  \textbf{Hit} \\
 & \textbf{(K)} & \textbf{WAF} & \textbf{WAF} & \textbf{(Byte/op.)} & \textbf{(Byte/op.)} & \textbf{Ratio} \\
\midrule
in-place & 229 & 2.00 & 2.36 & 1,858 & 4,378 & 0.932 \\
$\rightarrow$ out-of-place & 230 & 4.06 & 1.94 & 3,745 & 7,274 & 0.918 \\
$+$ comp. $+$ pagepack & 380 & 0.62 & 1.95 & 566 & 566 & 0.929 \\
$+$ GDT & 458 & 0.59 & 1.96 & 566 & 1,110 & 0.931 \\
$+$ NoWA & 510 & 0.60 & 1.07 & 567 & 606 & 0.931 \\
$+$ aligning GC Unit & 535 & 0.60 & 1.00 & 567 & 567 & 0.931 \\
$-$ comp. & 328 & 3.58 & 1.00 & 3,364 & 3,364 & 0.927 \\
\midrule[0.3pt]\bottomrule[1pt]
\end{tabular}
\end{table}

\module{Throughput.}
Applying all optimizations reduces total WAF and raises throughput from 229K OPS (in-place writes) to 535K.
Fewer logical writes reduce time spent waiting for I/O~\cite{DBLP:journals/pvldb/LeeAL23,DBLP:journals/pacmmod/WangQYZ25}, 
and lower SSD WAF shortens I/O latency~\cite{ssdiq}, further improving throughput.
Compression with page packing adds further gains by reducing read traffic per operation.

\module{DB WAF.}
For in-place, DB WAF is 2.00 due to double-write buffering.
Out-of-place writes remove this cost, but with the dataset occupying~90\% of SSD capacity and greedy DB GC, DB WAF rises to 4.06.
Compression lowers DB WAF to 0.62 by both giving DB GC additional OP space and reducing the user write volume itself.
GDT-based placement reduces it further to 0.59 by leveraging skew at GC time.
The NoWA pattern slightly increases DB WAF to 0.60 due to compensation writes, but, as shown next, the reduction in SSD WAF outweighs this cost.

\module{SSD WAF.}
In-place writes exhibit SSD WAF~=~2.36, a value that varies across models (see \cref{sec:varyingssds}).
Switching to append-per-zone out-of-place writes reduces this to 1.94, likely because the large sequential write pattern aligns better with SSD internals~\cite{DBLP:conf/fast/MinKCLE12}.
The NoWA pattern further reduces SSD WAF to 1.07.  
While the effect of applying NoWA or GC-unit alignment alone varies by runtime, device model, and workload skew, applying both together consistently eliminates SSD WA, achieving WAF 1.0.

\module{Logical \& physical writes per operation.}
Columns 5--6 of \cref{tab:drilldown} show logical (DB-issued) and physical (SSD-internal) bytes written per operation.
With all optimizations, logical writes drop by 3.27× mainly due to compression; without compression they increase to 1.81×.
Physical writes consistently decrease as SSD WAF approaches 1, eventually converging to logical writes.
Overall, the techniques substantially reduce physically executed writes.

\module{Buffer hit ratio.}
GC I/O can reduce buffer hit ratios because GC reads may evict cached pages.
The largest drop occurs when switching from in-place to out-of-place writes (93.2\%→91.8\%).
Compression improves the valid ratio and restores the hit ratio to 92.9\%; GDT increases it slightly to 93.1\%, since pages read during GC are more likely to be accessed soon (effectively acting as prefetching).
Overall, the optimizations have minimal impact on hit ratio.

\module{CPU overhead.}
In-place writes use 5\% CPU, while out-of-place writes with all optimizations use 8.3\%.
Most overhead comes from securing buffer frames for GC reads.
Since in-place writes already waste ~8\% CPU waiting for synchronous DWB, this extra usage is largely hidden.
Future work will reduce overhead by improving hash-table scans when locating clean frames.

\module{Memory usage.}
We kept the buffer pool size the same across in-place and out-of-place versions to match user writes per operation.
The out-of-place schemes require extra metadata (\cref{sec:impl}), which is kept in memory,
\revision[R2.D5]{amounting to up to 10.9~GB in the worst case when the device is fully utilized. This bound varies with system configuration, such as compression ratio and zone size.}
If memory is constrained, the system can keep only active metadata and spill the rest to disk, which we leave to future work.


\begin{figure}[t]
\includegraphics[clip, width=\columnwidth]{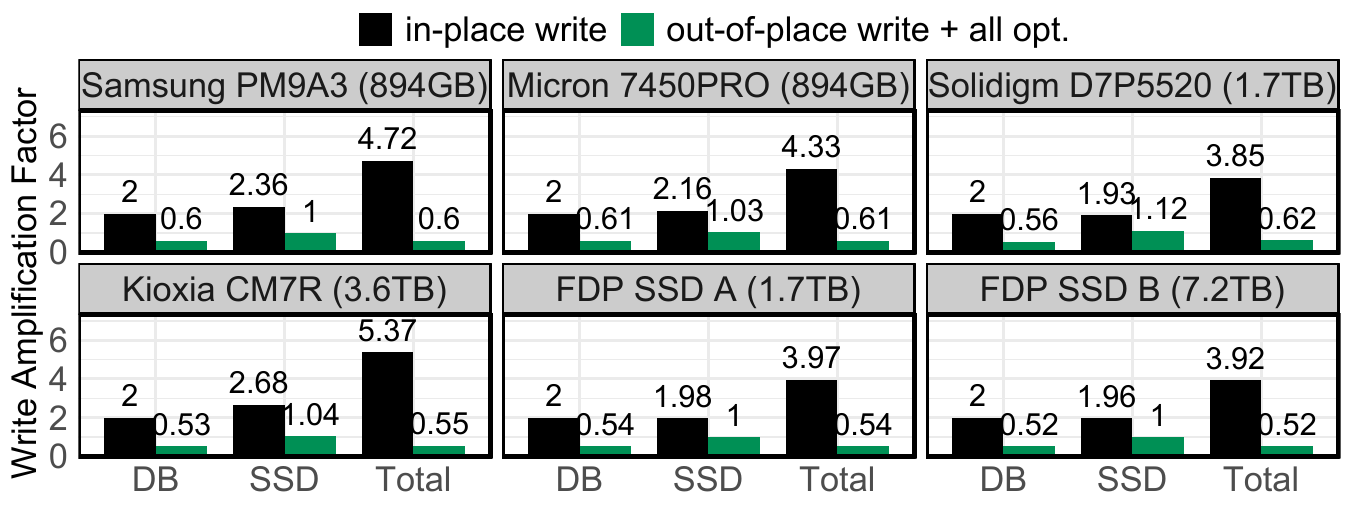}
\caption{\revision{WAF comparison across different SSD models (YCSB-A zipf $\theta$ = 0.8, SSD space 90\% full)}}
\label{fig:varyingssd}
\end{figure}

\subsection{Results Across Various Commodity SSDs}~\label{sec:varyingssds}

\module{Different commodity SSDs.}
To validate our optimizations across SSD models, we repeat the experiment on five additional enterprise SSDs, 
scaling datasets to ~90\% of each device’s capacity (YCSB-A, zipf $\theta$=0.8, 80~GB buffer pool).  
\cref{fig:varyingssd} reports DB, SSD, and total WAF.  
Out-of-place writes consistently reduce total WAF over in-place, with improvements ranging from 6.2× (Solidigm D7-P5520) to 9.76× (Kioxia CM7-R), 
demonstrating robustness across models, vendors, and capacities.  
Larger SSDs exhibit lower DB WAF due to having more OP space (e.g., 7.2\,TB → 752\,GB vs.\ 894\,GB → 94\,GB).  
SSD WAF under out-of-place remains near the optimal value of 1 for all devices except the Solidigm D7-P5520 (1.12), 
which inherently shows slightly higher WAF even on sequential workloads~\cite{ssdiq}.  
In contrast, SSD WAF under in-place ranges from 1.93 to 2.68.  
Overall, the optimizations reliably minimize SSD WAF across devices.


\begin{table}[t]
\begingroup
\setlength\doublerulesep{0.5pt}
\setlength{\aboverulesep}{0.2ex}
\setlength{\belowrulesep}{0.2ex}
\setlength{\tabcolsep}{0.12cm}
\renewcommand{\arraystretch}{0.95}
\small
\centering
\captionsetup{skip=6pt}
\caption{Throughput comparison on CNS and ZNS SSDs (YCSB-A zipf $\theta$ = 0.8; capacities: CNS = 894\,GB, ZNS = 951\,GB)}
\label{tab:zns1}
\begin{tabular}{l l l r|r}
\toprule[1pt]\midrule[0.3pt]
\textbf{} & \textbf{Configuration} & \textbf{Used} & \textbf{Logical DB} & \textbf{OPS} \\
\textbf{} & \textbf{} & \textbf{SSD} & \textbf{Size (GB)} & \textbf{(K)} \\
\midrule
\multirow{3}{*}{\shortstack{\textbf{in-place vs.}\\\textbf{out-of-place}}}
  & in-place             & CNS & 800  & 156.8 \\
  & oop + all opt.       & CNS & 800  & 322.2 \\
  & oop + comp. + GDT    & ZNS & 800  & 330.3 \\
\midrule
\multirow{3}{*}{\shortstack{\textbf{CNS vs. ZNS}\\\textbf{larger DB size}}}
  & oop + all opt.       & CNS & 1,500 & 250.7 \\
  & oop + comp. + GDT    & ZNS & 1,500 & 328.9 \\
  & oop + comp. + GDT    & ZNS & 1,596 & 277.5 \\
\midrule[0.3pt]\bottomrule[1pt]
\end{tabular}
\endgroup
\end{table}

\subsection{ZNS and FDP}\label{sec:znseval}

\module{Out-of-place writes benefit CNS, but even more so on ZNS.}
ZNS SSDs offer two key advantages over standard SSDs:  
(1) they naturally maintain SSD WAF~=~1 without requiring NoWA, and  
(2) they expose more usable capacity because no internal OP is needed.  
To evaluate these advantages, we run YCSB-A (zipf $\theta=0.8$) on a 951~GB ZNS SSD and an 894~GB CNS SSD with identical firmware.  
Since CNS cannot report physical writes (no OCP support), we compare throughput instead.  
With an 800~GB dataset (first group in \cref{tab:zns1}), CNS with out-of-place writes doubles the throughput of in-place writes, 
and ZNS further improves throughput---2.1× over CNS in-place and 1.03× over CNS out-of-place.

\module{ZNS vs. CNS with the same dataset size.}
To isolate the benefit of increased usable capacity on ZNS, we evaluate both devices with a 1{,}500~GB dataset (second group in \cref{tab:zns1}).
CNS in-place is omitted since the dataset exceeds its usable capacity.
ZNS outperforms CNS by 31\%, primarily due to lower DB WAF: the additional space reduces the valid ratio during DB GC and avoids NoWA-induced compensation writes.

\module{ZNS vs. CNS at the same fill factor.}
We also compare the devices at the same \emph{relative} fill.
Since 1{,}500~GB on an 894~GB CNS matches the fill factor of 1{,}596~GB on a 951~GB ZNS, we evaluate these configurations (final row in \cref{tab:zns1}).
Here, ZNS outperforms CNS by only $\sim$10\%, much smaller than the previous 31\% gain.
This indicates that most of the benefit stems from ZNS’s larger usable capacity, while removing NoWA overhead yields a smaller incremental gain.


\begin{table}[t]
\begingroup
\setlength\doublerulesep{0.5pt}
\setlength{\tabcolsep}{0.12cm}
\renewcommand{\arraystretch}{0.95}
\small
\centering
\captionsetup{skip=6pt}
\caption{\revision{Effect of FDP on WAFs and throughput (YCSB-A zipf $\theta$=0.8, 1.6\,TB dataset, FDP SSD A)}}
\label{tab:fdpresults}
\begin{tabular}{l l |rrrr}
\toprule[1pt]\midrule[0.3pt]
\textbf{Configuration} & \textbf{FDP} & \textbf{DB} & \textbf{SSD} & \textbf{Total} & \textbf{OPS} \\
\textbf{} & \textbf{Enabled?} & \textbf{WAF} & \textbf{WAF} & \textbf{WAF} & \textbf{(K)} \\
\midrule
in-place                  & disabled & 2.00 & 1.98 & 3.96 & 437 \\
oop + all opt.            & disabled & 0.57 & 1.00 & 0.57 & 541 \\
oop + all opt. -- NoWA    & enabled  & 0.54 & 1.00 & 0.54 & 553 \\
\midrule[0.3pt]\bottomrule[1pt]
\end{tabular}
\endgroup
\end{table}


\revision[R2.D9]{
\module{SSD WAF = 1 with NoWA vs.\ FDP.}
To isolate the effect of FDP, we evaluate three configurations:
(1) in-place writes;  
(2) out-of-place writes with all optimizations, using NoWA to ensure SSD WAF = 1; and  
(3) out-of-place writes with all optimizations except NoWA, using FDP placement hints to ensure SSD WAF = 1 instead.  
\cref{tab:fdpresults} reports DB WAF, SSD WAF, total WAF, and throughput for these configurations on YCSB-A (zipf $\theta=0.8$) using a 1.6~TB dataset.  
Both out-of-place configurations achieve SSD WAF = 1, whereas in-place writes show SSD WAF = 1.98.  
Without FDP, sustaining SSD WAF = 1 requires the NoWA pattern, whose compensation writes raise DB WAF to 0.57.  
With FDP enabled, placement hints prevent multiplexing directly and further reduce DB WAF to 0.54.  
As a result, total WAF decreases and throughput improves correspondingly.
}


\subsection{Effects of Workload Variation}\label{sec:differentworkload}

\begin{figure}[t]
\includegraphics[clip, width=\columnwidth]{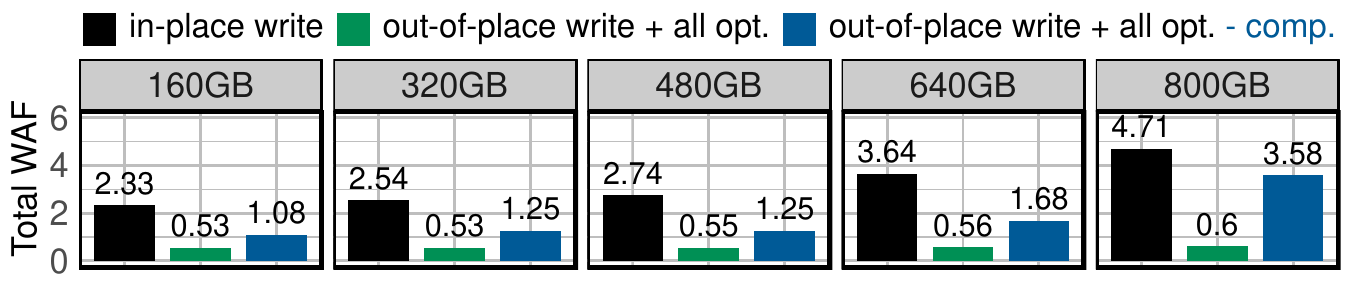}
\caption{Total WAF while varying dataset size (YCSB-A zipf $\theta$ = 0.8, Samsung PM9A3)}
\label{fig:dbsz}
\end{figure}

\module{Varying dataset size.}
We run YCSB-A (zipfian $\theta = 0.8$) with dataset sizes from 160~GB to 800~GB, setting the buffer pool to 10\% of each dataset.  
\cref{fig:dbsz} reports total WAF for in-place writes, out-of-place writes with all optimizations, and out-of-place writes without compression.  
With in-place writes, total WAF increases from 2.33 (160~GB) to 4.72 (800~GB), whereas out-of-place writes with all optimizations remain low (0.53-0.60).  
Without compression, WAF rises from 1.08 to 3.58 but still stays well below in-place writes.  
The benefits of our optimizations become more pronounced as datasets grow: in-place writes incur rising SSD WAF, while out-of-place writes maintain SSD WAF~=~1.  
Overall, our approach reduces both SSD and DB amplification, even without compression.

\begin{figure}[t]
    \includegraphics[clip, width=0.75\columnwidth]{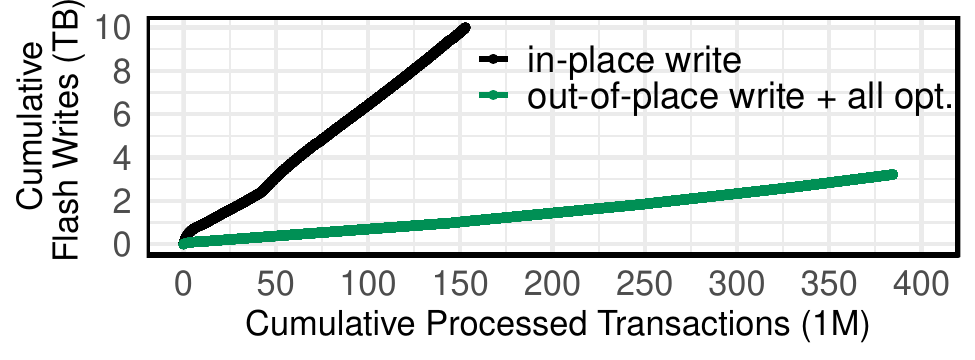}
    \caption{Comparison of cumulative flash writes per transaction during the same runtime (TPC-C 1.6\,TB, FDP SSD A)}
    \label{fig:tpcc}
\end{figure}

\module{Evaluating with TPC-C.}
Beyond YCSB-A, we run TPC-C~\cite{tpcc}, whose dataset grows as new orders arrive (unlike YCSB-A’s fixed size).
We load 15{,}000 warehouses on FDP SSD~A and run two-hour tests with in-place writes and with optimized out-of-place writes.
\cref{fig:tpcc} plots cumulative flash writes versus transactions.
In the same runtime, the optimized out-of-place configuration completes 2.45$\times$ more new-order transactions; for the same transaction count, in-place writes incur 7.2$\times$ more flash writes.
These results show our approach remains effective under a different workload.

%% file: related.tex
\section{Related Work}~\label{sec:related}

\module{Different ways to perform out-of-place writes.}
\revision[R3.W2]{ Out-of-place writes appear in several forms.
WiredTiger~\cite{wiredtiger-compaction} stores each table in a cyclic file and rewrites valid pages to the file head during compaction.
LSM engines~\cite{rocksdb-saf, leveldb, tigerbeetle-arch} also write out-of-place, reclaiming space only when compaction is triggered, making GC timing and reclaimed space hard to control.
In contrast, ZLeanStore reclaims space proactively at zone granularity~\cite{DBLP:conf/fast/LeeSHC15, nilfs2, lfs, zenfs, DBLP:conf/usenix/Hwang0KES25, skylight}, independent of other DBMS components~\cite{DBLP:journals/tos/LuPGAA17}.
Doing so at the DBMS layer allows GC to account for both DB- and SSD-level WA. }

\module{Optimizing data structures to reduce DB WA.}
A substantial amount of work has reduced WA by optimizing index structures.
B-tree-based storage engines such as WO-B-tree~\cite{WObtree}, Bf-tree~\cite{DBLP:journals/pvldb/HaoC24}, and B-epsilon tree~\cite{DBLP:journals/usenix-login/BenderFJJKPY015} buffer updates and apply them in batches to reduce index-induced WA.
LSM-tree engines were introduced to address this, and various works attempt to optimize it further~\cite{rocksdb-saf, leveldb, tigerbeetle_github, novelsm,DBLP:conf/edbt/VinconHRO0P18,LSMTreeCost_SIGMOD2024,Dayan_TODS2018,Liu2025_ArceKV,lsmdesign,Dayan_VLDB2022_Hybrid}.
These approaches are orthogonal to ours: they reduce WA at the index level, whereas our techniques reduce DB WA at the page level without changing the index structure.
Thus, they can be complementary when applied together.

\module{Reducing WA by mitigating space amplification.}
Write volume can be reduced by lowering overall storage consumption.
Techniques such as XMerge~\cite{DBLP:conf/cidr/AlhomssiL21} and B-tree node compaction~\cite{DBLP:journals/access/LeeAL23} 
improve node utilization and reduce disk writes; these index-level optimizations are orthogonal and can benefit from our approach.
\revision[R2.D4]{Disk-level compression in MySQL~\cite{mysql-comp}, WiredTiger~\cite{wiredtiger-compaction}, and RocksDB~\cite{rocksdb-saf} 
similarly reduces write volume and storage footprint, 
but is ineffective under 4~KiB page granularity, whereas we provide a solution for 4~KiB pages. }
Moreover, out-of-place--write engines do not examine how compression interacts with DB WAF induced by GC, whereas we explicitly analyze and address these effects.
Some SSDs also employ internal compression to increase overprovisioning and mitigate WA~\cite{DBLP:conf/osdi/ZuckTSH14}; 
our DB GC naturally benefits from compression as well, but, unlike SSDs, we can flexibly enable, disable, or tune compression based on application needs.

\module{Mitigating WA upon GC.}
Prior work proposes a wide range of GC algorithms that classify pages by lifetime, recency, or spatial locality to reduce WA during GC~\cite{DBLP:conf/eurosys/ZhouWHHXZ15, DBLP:journals/corr/DayanB15, DBLP:conf/fast/MinKCLE12, DBLP:journals/cacm/LangeNY23, DBLP:journals/tos/Desnoyers14, DBLP:conf/fast/LeeSHC15, lockv, DBLP:journals/tos/LuPGAA17, DBLP:journals/pvldb/YuMKLMK22, DBLP:journals/pe/Houdt14, DBLP:conf/hotstorage/ShafaeiDF16, lomet2020efficientlyreclaimingspacelog}.
Our approach differs in that we predict deathtimes of individual pages using database-specific semantics, 
which lower layers such as SSD GC cannot exploit~\cite{DBLP:journals/cacm/LangeNY23,DBLP:conf/eurosys/HeKAA17, DBLP:journals/pomacs/LangeNY25}.
To our knowledge, this is the first work to integrate GDT-based GC within a DBMS, leveraging workload knowledge that is only available to the DBMS.

\module{Optimizing system write patterns for SSDs.}
Prior work has examined how applications can tailor I/O to SSD characteristics to exploit internal parallelism~\cite{DBLP:conf/eurosys/HeKAA17, DBLP:journals/pvldb/KakaraparthyPPK19, DBLP:journals/pvldb/HeXLJZYML23}, analyzed tree-based access patterns~\cite{DBLP:journals/pvldb/DidonaISK20}, and characterized SSD behavior under diverse workloads~\cite{DBLP:journals/tos/YadgarGJS21, DBLP:conf/hotstorage/YadgarG16}.
Although these studies improve latency or throughput, they do not explicitly target SSD WAF; even when WAF is considered, it is not fully eliminated~\cite{DBLP:conf/osdi/YangPGTS14, DBLP:conf/fast/Curtis-MauryKRM24}.
In contrast, our DBMS GC directly minimizes SSD WAF while accounting for its interaction with DB WAF to reduce total WAF.

\module{Zoned storage for ZNS.}
Several systems~\cite{ssdfs, zenfs, DBLP:conf/usenix/Hwang0KES25, yang2024zonedstorageoptimizedflash, DBLP:conf/hotstorage/LeeLL022, DBLP:journals/fgcs/HaS25} 
leverage ZNS SSDs~\cite{DBLP:conf/usenix/BjorlingAHRMGA21} to exploit the SSD WAF of~1.
While effective, these designs are tied to ZNS hardware and thus limited to environments where such devices are available.
In contrast, our approach treats ZNS as an optional optimization rather than a requirement, making it applicable across a broader range of devices.

\module{Systems using FDP.}
The FDP interface was introduced only recently~\cite{nvme_fdp_tp4146}, and emerging work explores its potential for reducing WAF~\cite{DBLP:conf/hotstorage/ZhangPCK23,eurosys25,george2024fdp}.
Our work identifies the optimal use case for FDP placement hints from the perspective of out-of-place DBMSs to achieve an SSD WAF of~1.
We further suggest appropriate settings for DB GC granularity and the number of active zones under FDP.

%% file: conclusion.tex
\section{Conclusion}\label{sec:conclusion}
This paper demonstrates that co-designing a DBMS and an SSD can improve throughput and extend SSD lifespan.
Our four out-of-place schemes reshape the DBMS write pattern to mitigate WA at both the DB and SSD layers. 
We further show that the design naturally leverages ZNS and FDP.
\revision[R1.W1]{Future work includes extending these techniques to LSM engines, supporting multi- or shared-device deployments, and exploring disk- and HM-SMR--based systems.}

Beyond DBMSs and SSDs, our findings apply broadly to storage systems.
Our large sequential write patterns also benefit disk-based systems~\cite{DBLP:journals/corr/abs-2205-11753,DBLP:conf/sosp/BornholtJACKMSS21,skylight}.
Systems with their own translation layer and garbage collection can adopt our DB-level WAF optimizations, 
while log-structured filesystems can use the NoWA pattern to reduce aging~\cite{DBLP:conf/hotstorage/ParkE22,DBLP:conf/usenix/KadekodiNG18}.
We hope this work helps storage designers better understand the implications of system writes for devices and how to fully exploit device capabilities.

%% file: appendix.tex
\section{Frequently Asked Questions}\label{sec:appendix}
In this appendix, we provide answers to a set of plausible questions that may arise from our paper.

\subsection{Should the DBMS really do all this work?}\label{sec:q1}

\module{Conventional layer responsibilities.}
Traditionally, DBMSs, file systems, and SSDs are designed to be responsible for different tasks.
In terms of storing data, databases first interact with clients to generate data in their preferred manner.
Then, filesystems communicate with the underlying device to store it on SSDs.
Finally, SSDs ensure that data persist on the physical NAND flash chips.
To effectively perform these tasks, each component has evolved significantly, as studied within its respective research community.

\module{WA manifests across layers in a top-down manner.}
Although these three layers handle conceptually orthogonal responsibilities for persisting data on non-volatile storage devices, 
they influence one another in practice---primarily in a top-down manner. 
The DBMS issues writes (possibly through the filesystem), while the SSD passively receives them. 
The stream of write requests is transformed multiple times before it physically lands on the flash chips, 
depending on whether the DBMS, filesystem, and SSDs implement out-of-place writes.
Consequently, how WA manifests across layers depends on how the DBMS issues writes in the first place.

\module{WA across layers is interdependent: a simple example.}
Consider the simple setup: an out-of-place-write DBMS writes directly to an SSD and stores a logical dataset of around half of the SSD capacity.
Two forms of GC then operate: DB GC and SSD GC.
The effective physical dataset size is determined by the point at which DB-level GC reclaims invalidated space.
This, in turn, dictates how the available OP space is split between DB-level and SSD-level GC, as illustrated below:
\begin{center}
\includegraphics[clip, width=0.9\columnwidth]{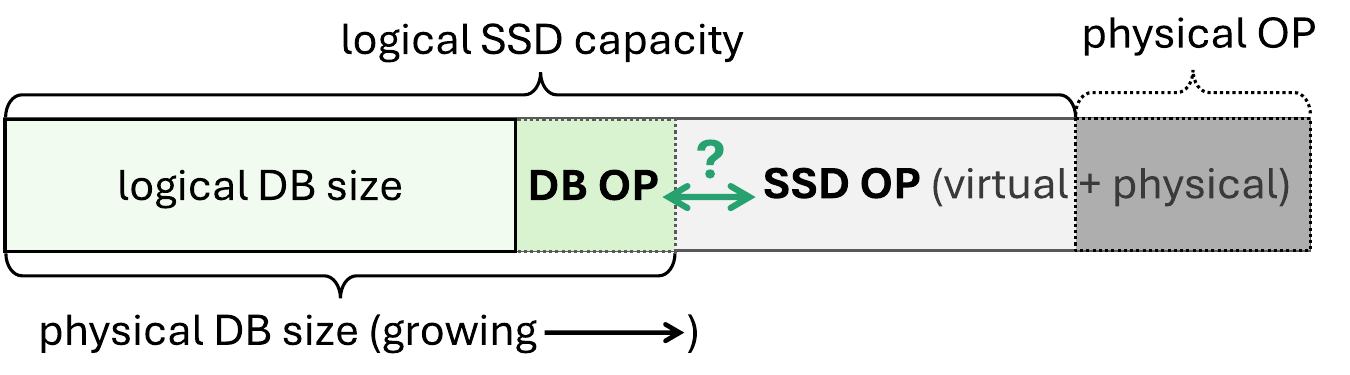}
\end{center}

\module{Why total WAF matters.}
Based on this OP split, DB WAF and SSD WAF vary inversely as they contend for OP space, and total WAF varies accordingly.
As shown below, as the physical dataset size increases, DB WAF goes down but SSD WAF goes up:
\begin{center}
\includegraphics[clip, width=0.6\columnwidth]{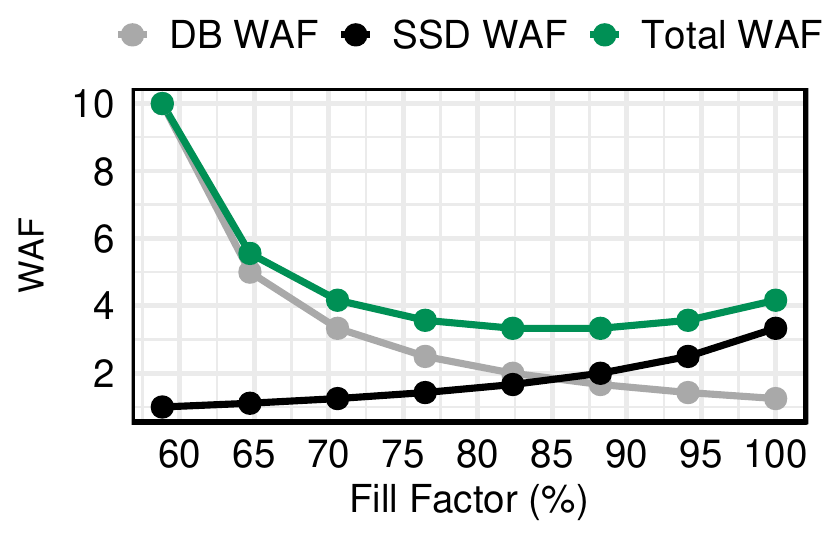}
\end{center}
Now, if we aim solely to minimize DB WAF, then DB GC will use the full logical SSD capacity (100\% on the x-axis). 
This indeed keeps DB WAF relatively low, around 1.3.
However, because the SSD was then deprived of the OP space, the SSD WAF increases to approximately 4.2 in return. 
As a result, even though we intended to reduce WA, the SSD ends up performing 5.46× (1.3 × 4.2) more internal writes. 
This total WAF is worse than when the DB GC is triggered earlier, at an 85\% fill factor, for instance, where the total WAF is 3.3. 
This outcome is counterintuitive from the DBMS perspective, but illustrates what happens when the SSD’s internal behavior is ignored.

\module{The case for total WAF consideration.}
This motivates the total WAF formulation: focusing on individual WAFs is counterproductive when reducing one worsens another. 
Thus, total WAF (DB WAF × SSD WAF) better reflects the true cost by representing the total flash writes per transaction. 
We argue that considering total WAF is a key contribution of this work, 
whereas prior studies typically examine WA only within individual layers. 
WA should therefore be analyzed across layers---i.e., as total WAF---rather than in isolation.

\module{Why DBMS, and not SSDs or filesystems?}
To effectively address total WAF, we argue that the DBMS---rather than the filesystem or the SSD---should take the lead.
This may sound counterintuitive, as filesystems and SSDs sit closer to the lower layers and therefore seem better positioned to optimize it.
Moreover, the storage community has made relentless efforts to minimize SSD WAF, and it remains an active research topic today.
However, despite this radical shift, we claim that the DBMS is more suited to, and therefore should, optimize SSD writes for two main reasons.
First, despite extensive research, our prior work shows that most enterprise SSDs still exhibit surprisingly high WAF~\cite{ssdiq}.
Second, the DBMS has the greatest visibility into workload behavior, whereas filesystems and commodity SSDs have virtually none.

\module{Modern SSDs still exhibit high WAF.}
Our prior research~\cite{ssdiq} has shown that, surprisingly, standard commodity SSDs are not capable of interpreting workload characteristics, even for the simplest skewed patterns. 
As a result, they inevitably suffer from high SSD WAF.
As shown in Figure~3 of our prior work~\cite{ssdiq}, we evaluated a two-zone workload by splitting space 90:10\% and using an access pattern of 10:90\%.
Different enterprise SSDs from multiple vendors all exhibited consistently high WAF, ranging from 3 to 4.8, and their WAF trends closely matched the behavior of greedy GC when compared with simulator results.
This clearly suggests that they do not implement intelligent GC algorithms that exploit workload characteristics.

\module{SSDs lack workload knowledge to reduce WA.}
The fundamental reason why it is intrinsically difficult to build intelligent GC inside SSDs is their lack of visibility into the workload~\cite{DBLP:journals/pomacs/LangeNY25}.
Extensive research has attempted to infer workload characteristics inside SSDs to improve GC~\cite{DBLP:journals/pomacs/LangeNY25,DBLP:journals/cacm/LangeNY23,DBLP:journals/pvldb/KangCOL20},
but such inference introduces significant overhead and remains inherently imperfect.
This is one of the key motivations behind proposals such as open-channel SSDs, ZNS, open-programmable SSDs~\cite{DBLP:conf/micro/ParkLLBSCC24}, and newer interfaces like multi-stream and FDP.
These interfaces encourage the application to manage the SSD in a workload-aware manner, or at least share valuable workload information with the device~\cite{SNIA_SDC_2025_TotalCostAndPerformanceOfSSDs, SNIA_SDC_2025_OptimizingHyperscaleFlashStorage}.
This indicates that, without some form of host-level coordination---such as what this paper proposes---SSDs alone cannot effectively address this problem.

\module{Filesystems cannot address total WAF.}
Filesystems are also ill-suited for reducing total WAF.
Situated between the DBMS and the SSD, they lack visibility into database-level access patterns and have no control over SSD-internal data placement.
Consequently, they possess neither the workload knowledge required to lower DB WAF nor the low-level interface needed to mitigate SSD WAF~\cite{DBLP:conf/usenix/BjorlingAHRMGA21}.
These limitations are evident even in filesystems explicitly designed to reduce SSD WAF.
For example, prior work shows that RocksDB running on ZenFS consistently achieves lower SSD WAF than RocksDB on F2FS with ZNS support~\cite{DBLP:conf/usenix/BjorlingAHRMGA21}.
Similarly, with FDP-enabled SSDs, SSD WAF is higher when XFS is used, compared to when RocksDB issues FDP placement hints directly without XFS~\cite{fdprocksdb}.
In both cases, filesystem mediation worsens WAF despite device-aware extensions, demonstrating that filesystems cannot match the effectiveness of DBMS-level write control.
Moreover, when the DBMS already employs out-of-place writes, the filesystem contributes little beyond adding an additional layer of indirection and system-call overhead~\cite{DBLP:conf/icde/NguyenL24,DBLP:conf/icit/HinesCA23,DBLP:journals/corr/SearsIG07,DBLP:journals/pvldb/GaffneyPBHKP22}.
While operating systems provide useful abstractions, these features offer limited benefit to a DBMS that already manages its own memory and storage layout.

\module{DBMS, at the highest layer, can mitigate total WAF the most.}
Therefore, we argue that the DBMS is the most suitable layer for mitigating total WAF.
Although this represents a more unconventional departure from the traditional DB-filesystem-SSD three-layer architecture,
the DBMS benefits from operating at the highest layer.
It possesses the richest knowledge of workload behavior at the topmost layer.
Informed by schema structure, access patterns, and query semantics, DBMS GC has far more potential to outperform SSD GC, and thus mitigate total WAF.

\module{Out-of-place writes are inevitable to reduce WA.}
To achieve this, switching from in-place to out-of-place writes becomes necessary; without this flexibility, the DBMS cannot tailor writes to workload characteristics.
Once switched to out-of-place writes, the DBMS gains full control over how and when writes are issued,
allowing it to coordinate write patterns with awareness of both application semantics and SSD behavior.
While adopting out-of-place writes does introduce architectural implications for other DBMS components such as logging and recovery~\cite{WObtree},
we claim it is the only way to enable the SSD write optimizations proposed in this paper.
\subsection{When should the DBMS care about WAF?}\label{sec:q2}
\noindent High SSD WAF becomes critical to SSD performance and lifespan under typical OLTP workloads, 
which are characterized by growing dataset sizes (i.e., varying fill factor), high write intensity, and skewed access patterns~\cite{ssdiq, DBLP:journals/pvldb/KangCOL20, Gray1993DebitCredit}.
Below, we explain how each factor influences SSD WAF and throughput, and how our optimizations can affect them.

\module{Case 1: high fill factor.}
First, with respect to fill factor, SSD WAF increases as the SSD becomes more fully occupied.
This behavior explains why the effectiveness of our optimizations depends on the logical dataset size, as shown in \cref{fig:drilldown:a}.
As the dataset occupies a larger fraction of the SSD capacity, SSD WAF naturally increases for in-place-write LeanStore, 
whereas our SSD-level optimizations guarantee SSD WAF~$=1$ regardless of dataset size.
One way to mitigate this effect is to provision substantially more SSD capacity per logical dataset, 
effectively operating the SSD at a lower fill factor by amortizing space cost.
However, under high fill factors, the system inevitably suffers from elevated SSD WAF.

\module{Case 2: high write intensity workload.}
Second, when the DBMS issues writes at a high rate, SSD throughput degrades substantially under high SSD WAF.
Under high write intensity, flash read and program operations are more likely to arrive on busy flash LUNs \cite{lerner2024principles}.
This increases contention and amplifies the performance penalty of elevated WAF.
Furthermore, high SSD WAF directly accelerates wear, shortening the SSD’s lifetime.
Thus, the importance of reducing SSD WAF is significantly heightened in high write-intensity conditions.
In this scenario, reducing DB WAF becomes increasingly beneficial, as it lowers the logical write volume to the SSD, 
effectively reducing the write intensity, potentially by half via compression, for instance.

\module{Case 3: skewed write access pattern.}
Finally, SSD WAF can remain high even under explicitly hot/cold--separated workloads.
In our prior study~\cite{ssdiq}, we showed that enterprise SSDs from multiple vendors exhibit \emph{higher} WAF under skewed workloads than under uniform random access.
Figures~3 and~4 in our previous paper~\cite{ssdiq} demonstrate that these SSDs generally lack sophisticated GC mechanisms and therefore incur high WAF unless the hot data fits entirely within their internal write buffers.
As we argue in \cref{sec:GDT}, it is therefore necessary to implement workload-aware GC mechanisms, such as GDT-based GC, at the application layer, since commodity SSDs rarely exploit such workload characteristics effectively.
Thus, if an in-place DBMS generates an extremely skewed write access pattern, it is likely to suffer from increased SSD WAF.
In contrast, DBMSs that perform out-of-place writes exhibit no such correlation, because frequently accessed PIDs are always written to different locations.

\subsection{What if disks are used?}\label{sec:q3}
\noindent \module{Out-of-place writes remain beneficial on disks.}
When running ZLeanStore on traditional hard disks, we expect it to still benefit from the large sequential writes enabled by out-of-place writes.
Additionally, since we propose not only SSD WAF optimizations but also DB WAF optimizations,
ZLeanStore will outperform in-place-write LeanStore, as it reduces logical writes to the device through compression (\cref{sec:comp}) and GDT-based GC (\cref{sec:GDT}).
Furthermore, HM-SMRs are also zoned storage devices; therefore, ZLeanStore can run on top of them with zone-command support (\cref{sec:zns}).
Hence, even with traditional disks, it is beneficial for DBMSs to adopt out-of-place writes (which are mandatory for HM-SMRs) and apply some of our optimizations accordingly.

\subsection{What happens if a filesystem is used?}\label{sec:q4}
\module{Removing the filesystem is no longer unusual.}
In this paper, we assume that the DBMS issues read and write requests directly to the block device, without an intervening filesystem (\cref{fig:oop2}).
Before discussing the implications of introducing a filesystem, it is important to note that writing directly to the block device is not new.
Recent work has shown that filesystems may slow down transaction throughput due to system call overhead
~\cite{DBLP:conf/icde/NguyenL24,DBLP:conf/icit/HinesCA23,DBLP:journals/corr/SearsIG07,DBLP:journals/pvldb/GaffneyPBHKP22, DBLP:journals/tos/AghayevWKNGA20},
especially for small objects typical of OLTP workloads.
Consequently, several research efforts pursue DB--OS co-design by eliminating the filesystem layer entirely when running DBMSs~\cite{DBLP:journals/pvldb/SkiadopoulosLKK21}. 
Accordingly, it is no longer unusual for databases to run directly on top of the block device.

\module{Total WAF when using a filesystem.}
Nevertheless, the traditional configuration places the DBMS above a filesystem rather than directly on the block device.
To evaluate the effect of filesystems on SSD WAF, we conduct separate experiments by running the in-place-write baseline, LeanStore, on top of three different filesystems: ext4, F2FS, and XFS.
We run the same benchmark as in \cref{fig:intro} and report DB WAF, filesystem WAF, SSD WAF, total WAF, and throughput:

\begin{table}[h]
\centering
\footnotesize
\begin{tabular}{l|rrrrr}
\toprule[1pt]\midrule[0.3pt]
\textbf{Filesystem (LeanStore config.)} & \textbf{DB} & \textbf{FS} & \textbf{SSD} & \textbf{Total} & \textbf{OPS} \\
\textbf{} & \textbf{WAF} & \textbf{WAF} & \textbf{WAF} & \textbf{WAF} & \textbf{(K)} \\
\midrule
ext4 (in-place)                         & 2.00 & 1.00 & 2.41 & 4.87 & 212 \\
F2FS (in-place)                         & 2.00 & 1.03 & 2.48 & 5.13 & 199 \\
XFS (in-place)                          & 2.00 & 1.00 & 2.41 & 4.83 & 212 \\
-- (in-place)                           & 2.00 & --    & 2.36 & 4.72 & 237 \\
-- (out-of-place + all opt.)            & 0.62 & --    & 1.00 & 0.62 & 470 \\
-- (out-of-place + all opt.\ $-$\ comp.)& 3.58 & --    & 1.00 & 3.58 & 328 \\
\midrule[0.3pt]\bottomrule[1pt]
\end{tabular}
\end{table}
\noindent
As shown in the table, the filesystem does not improve WAF; in some cases, it even worsens it and reduces throughput due to the additional layer.
Even F2FS, which is widely regarded as flash-friendly, does not address SSD WAF.

\module{Filesystem should participate in mitigating total WAF.}
If a filesystem is used, the filesystem must share responsibility for mitigating total WAF alongside the DBMS.
The filesystem should be the component performing out-of-place writes, as in log-structured filesystems~\cite{DBLP:conf/fast/LeeSHC15, nilfs2, lfs}, and should align its segment size with the SSD’s GC unit.
For in-place DBMSs, the filesystem alone is responsible for mitigating SSD WAF; DB WAF remains unaddressed.
When the DBMS also performs out-of-place writes, DB GC must account for the filesystem’s write patterns when selecting victim zones, after aligning the DBMS’s GC unit with the filesystem’s segment size.

\subsection{What about shared/multi-device scenarios?}\label{sec:q5}
We assume a single-instance, single-device setup in this work.
Oftentimes, especially in cloud environments, storage is shared across instances or spans multiple devices.
Although we do not explore these settings, our optimizations can potentially be applied to such scenarios.
We leave a detailed exploration of shared-device and multi-device environments for future work.

\module{Shared device scenario.}
When multiple database instances share a physical device through a virtualization layer, DB-level WAF mitigation remains the responsibility of each DBMS space manager within the logical capacity exposed by the virtual disk (e.g., VMDK).
The DBMS observes and manages only this guest-visible logical namespace and can track how much logical space it has allocated or left free~\cite{vmware-vsphere-vmfs}.
However, it has no visibility into how these logical allocations map to physical offsets in the datastore or how much physical space is actually consumed.
Consequently, DB-level WAF and SSD-level WAF are addressed at different layers of the stack.
At the DB level, the space manager must decide whether to consume remaining logical free space (implicitly relying on physical headroom) or to reclaim already used space earlier to control DB-level WAF.
SSD-level WA mitigation, in contrast, requires coordination below the guest boundary.
The virtualization layer’s storage subsystem (e.g., the VMFS layer managing VMDKs) could be extended to support out-of-place writes across VMs, maintain a global view of write streams issued by different tenants, and allocate physical space according to our proposed optimizations, including aligning GC units (\cref{sec:gcgranularity}) and adopting the NoWA pattern (\cref{sec:nowa}).
Additionally, if the hypervisor and device stack expose FDP to guests, FDP hints could be propagated through the virtualization layer to the device~\cite{SNIA_SDC_2025_SSDVirtualization}.

\module{Multi-device scenario.}
When multiple SSDs are used and the DBMS directly manages them, the space manager should coordinate usable space and garbage collection across devices to optimize DB-level WAF.
This configuration gives the DBMS greater flexibility, since garbage collection can be managed independently on each device while balancing logical utilization across them.
For SSD-level WAF optimizations, the device manager should locally track the active-group history on each device to preserve the NoWA pattern.
If multiple SSDs are abstracted as a single logical block device by RAID or an array controller, the DBMS no longer retains ownership of physical placement decisions, and coordination across devices is therefore handled by the storage subsystem.
For instance, in multi-SSD storage systems such as ONTAP~\cite{DBLP:conf/fast/Curtis-MauryKRM24} and FlashArray~\cite{PureStorageFlashArrayArchitecture}, a centralized storage-controller placement and space-management layer (e.g., WAFL in ONTAP) is the natural layer to implement our SSD-level WAF optimizations.

\subsection{Is doublewrite buffering really necessary?}\label{sec:q6}
In \cref{sec:oop}, we use the avoidance of doublewrite buffering as one motivation for advocating out-of-place writes instead of in-place updates.
We address three main aspects of doublewrite buffering.

\module{DWB is enabled by default.}
First, doublewrite buffering is enabled by default in InnoDB to ensure data integrity.
According to the MySQL reference manual, it should only be disabled when performance is prioritized over durability~\cite{mysql-dwb}.
Without it, InnoDB cannot recover from torn-page failures caused by power loss during a page write.
Although a checksum mismatch enables corruption detection, recovery requires an intact page copy.
For this reason, we use doublewrite buffering as the baseline.

\module{DWB exists across many systems.}
Second, this mechanism is not unique to MySQL.
Many systems employ similar techniques to guard against torn writes.
PostgreSQL, for example, logs a full-page image on the first modification after each checkpoint (\emph{full-page writes})~\cite{postgresql/doc/dwb},
and SQLite preserves the old page in its rollback journal~\cite{sqlite-journal}.
Other systems, such as CedarDB~\cite{cedardb-dwb}, XtraDB~\cite{xtradb-dwb}, and MariaDB~\cite{mariadb-dwb}, also use doublewrite buffering.
By contrast, SQL Server~\cite{sqlserver-pageverify} and Oracle~\cite{oracle-block-checksum} detect but cannot automatically recover from torn-page writes.
Even AWS provides page-level atomic writes on EC2 instances to allow doublewrite buffering to be safely disabled~\cite{aws-dwb}.

\module{Hardware/OS may not guarantee atomicity.}
Finally, hardware-level atomicity is insufficient.
Even when bypassing the filesystem and writing directly to NVMe devices,
atomicity is guaranteed only at the sector level (typically 512~B, as in most SSDs used in our experiments, or 4~KiB)~\cite{nvmexpress-spec-1.1}.
Database pages are much larger (e.g., 16~KiB in MySQL and 8~KiB in PostgreSQL),
so torn pages remain possible.
Moreover, PostgreSQL relies on the OS page cache (i.e., buffered I/O by default), so atomicity also depends on the filesystem implementation.

All of these concerns are eliminated once the system adopts out-of-place writes.

\subsection{How can we calculate DB WAF on other DBMSs?}\label{sec:q8}
In the total WAF formulation, SSD WAF is relatively straightforward to obtain by comparing the physical flash writes reported via OCP commands with the logical writes issued by the DBMS to the SSD (if the SSD supports them).
In contrast, computing DB WAF is often less clear, as it is nontrivial to distinguish necessary user writes from redundant writes.
In this paper, we define user writes as the sum of eviction writes and checkpoint writes, assuming that the maximum WAL file size equals the buffer pool size.
Different DBMSs implement checkpointing differently and may incur additional writes due to system-specific mechanisms (e.g., RocksDB compaction) or writing strategies (e.g., in-place versus out-of-place updates).
In this section, we describe how DB WAF is calculated for four systems: MySQL/InnoDB, PostgreSQL, WiredTiger, and RocksDB.
We assume that all WAL logs are stored on a separate device, consistent with \cref{fig:intro}.

\module{DB WAF of MySQL/InnoDB.}
As mentioned and evaluated in \cref{sec:intro}, MySQL’s doublewrite mechanism largely determines the DB WAF, similar to in-place-write systems such as the in-place LeanStore baseline.
Before flushing dirty pages from the LRU list or the flush list in the buffer pool, InnoDB first copies each full page into the DWB~\cite{mysql-dwb}.
The pages are then written sequentially to the DWB area on disk and subsequently written again to their final locations in the tablespace.
The size of the DWB is configured by the number of doublewrite files and the number of pages per doublewrite batch (e.g., \texttt{innodb\_doublewrite\_files} and \texttt{innodb\_doublewrite\_pages}).
This space is allocated per buffer pool instance, resulting in only a few megabytes of doublewrite space per instance under default settings~\cite{mysql-sourcecode}.
If the same page is selected for flushing multiple times before the DWB itself is persisted, InnoDB coalesces these requests and retains only the most recent page image in the DWB, avoiding redundant writes within a single doublewrite batch.
Furthermore, with respect to checkpoint writes, InnoDB exposes the maximum redo log size as a configurable parameter~\cite{mysql-checkpoint}.
When this parameter is set sufficiently large, checkpointing is no longer forced by log-space pressure and thus avoids unnecessary checkpoint-induced page flushes.
Consequently, the DB WAF can be expressed as follows:
\[
\text{WAF}_{\text{MySQL}} =
\frac{\text{LRU list flushes} + \text{flush list flushes} + \text{DWB}}
     {\text{LRU list flushes} + \text{flush list flushes}}
\]

\module{DB WAF of PostgreSQL.}
Unlike MySQL, PostgreSQL implements doublewrite protection via full-page writes to the WAL rather than a separate doublewrite file~\cite{postgresql/doc/dwb},
which amplifies WAL traffic and increases checkpoint frequency.
This behavior is particularly harmful for OLTP workloads with small updates, where the amount of data modified per page flush is typically much smaller than the page size~\cite{nvppl}.
As full-page writes indirectly amplify user writes, we derive the DB WAF for PostgreSQL as follows.
First, we run PostgreSQL with full-page writes disabled and tune the checkpoint frequency such that the WAL size remains proportional to the buffer pool size, since the maximum WAL size is not strictly enforced.
Under this configuration, we measure the ratio of eviction writes to checkpoint writes and infer the amount of checkpointing required to truncate the WAL.
Second, we enable full-page writes and measure eviction and checkpoint writes.
Using the previously obtained ratio, we estimate the checkpointing required for the observed eviction writes and attribute the remaining checkpoint writes to full-page writes.
We define the DB WAF for PostgreSQL as follows:
\[
\text{WAF}_{\text{Postgres}} =
\frac{\text{eviction} + \text{checkpointing (including full-page)}}
     {\text{eviction} + \text{required checkpointing}}
\]
Here, we exclude vacuuming- and pruning-induced writes~\cite{PostgreSQLDoc}, as PostgreSQL’s out-of-place version updates primarily incur index-level WA rather than page-level WA.

\module{DB WAF of WiredTiger.}
For out-of-place-write systems, DB WAF primarily arises from space reclamation rather than doublewrite buffering.
In WiredTiger, checkpoints identify unreferenced pages and return their blocks to the free list~\cite{WiredTiger_Checkpoint_Arch}.
If space reclamation proceeds too slowly, data files may grow without bound and eventually exhaust available storage~\cite{MongoDB_WiredTiger_DataSize_Increase_2021}.
Although manual invocation of \texttt{compact()} can reclaim space by rewriting live pages and truncating files~\cite{wiredtiger-compaction}, our separate experiments show that this incurs substantial overhead.
We define DB WAF for WiredTiger in terms of space reclamation overhead.
Let required checkpointing be the minimum checkpoint-induced write volume under a dirty-page-to-log-write ratio with a buffer-pool-sized WAL.
Any checkpointing beyond this minimum is attributed to space reclamation and considered DB WAF.
If \texttt{compact()} is used, its write volume can be counted as space reclamation overhead in place of the excess checkpoint writes.
Accordingly, we define the DB WAF for WiredTiger as:
\[
\text{WAF}_{\text{WiredTiger}} =
\frac{\text{eviction} + \text{checkpointing} + \text{compact}}
     {\text{eviction} + \text{required checkpointing}}
\]

\module{DB WAF of RocksDB.}
In RocksDB, DB WAF arises from LSM-tree compaction.
Each update is written to both the memtable and the WAL, and when the active memtable exceeds its flush threshold, it becomes immutable and is flushed to a level-0 SST file. 
Whether that flush triggers a cascade of compaction writes depends on the compaction strategy and thresholds such as the L0 file-count trigger, level size targets, and target SST file sizes~\cite{rocksdb_db_bench}.
As a result, LSM-tree-driven DB WAF varies significantly with such compaction policies and parameters, as extensively studied in prior work~\cite{LSMTreeCost_SIGMOD2024, Dayan_TODS2018, Liu2025_ArceKV, lsmdesign, Dayan_VLDB2022_Hybrid}.
Therefore, these configurations should be tuned to workload characteristics. 
It is important to note that RocksDB is not page-based, unlike the three systems above.
Thus, its DB WAF is not directly comparable to theirs.
Nevertheless, we define the DB WAF for RocksDB as follows:
\[
\text{WAF}_{\text{RocksDB}} =
\frac{\text{memtable flush} + \text{compaction}}
     {\text{memtable flush}}
\]